\documentclass[fleqn,usenatbib]{mnras}
\pdfoutput=1

\usepackage[T1]{fontenc}
\usepackage{ae,aecompl}
\usepackage{color, soul}
\usepackage{verbatim}
\usepackage{units}
\usepackage{subfig}
\usepackage{pdflscape}
\usepackage{svg}
\usepackage{bm}
\usepackage{stfloats}
\usepackage{float}
\usepackage{multicol}
\usepackage{hyperref}
\usepackage{multirow}

\usepackage{graphicx}	
\usepackage{amsmath}	
\usepackage{amssymb}	
\usepackage{epstopdf}
\usepackage[normalem]{ulem} 

\newcommand\clearrow{\global\let\rowmac\relax}
\newcommand{\cmmnt}[1]{\ignorespaces}

\title[Hypervelocity stars from the LMC and MW]{Comparing hypervelocity star populations from the Large Magellanic Cloud and the Milky Way}

\author[Evans et al.]{
F. A. Evans$^{1}$\thanks{E-mail: evans@strw.leidenuniv.nl},
T. Marchetti$^{2}$,
E. M. Rossi$^{1}$, J. F. W. Baggen$^1$, S. Bloot$^{1}$, \\
$^{1}$Leiden Observatory, Leiden University, PO Box 9513, NL-2300 RA Leiden, The Netherlands\\
$^{2}$European Southern Observatory, Karl-Schwarzschild-Strasse 2, 85748 Garching bei M{\"u}nchen, Germany \\
}

\date{Accepted XXX. Received YYY; in original form ZZZ}

\pubyear{2021}

\begin{document}
\label{firstpage}
\pagerange{\pageref{firstpage}--\pageref{lastpage}}
\maketitle

\begin{abstract}
 We predict and compare the distributions and properties of hyper-velocity stars (HVSs) ejected from the centres of the Milky Way (MW) and the Large Magellanic Cloud (LMC). In our model, HVSs are ejected at a constant rate -- equal in both galaxies -- via the Hills mechanism and are propagated in a combined potential, where the LMC orbits the MW on its first infall. By selecting $m>2\, \mathrm{M_\odot}$ HVSs well-separated from the Magellanic Clouds and Galactic midplane, we identify mock HVSs which would stand out from ordinary stars in the stellar halo in future data releases from the \textit{Gaia} satellite and the Vera C. Rubin Observatory's Legacy Survey of Space and Time (LSST). We find that in these deep surveys, LMC HVSs will outnumber MW ones by a factor $\sim 2.5$, as HVSs can more easily escape from the shallower potential of the LMC. \textcolor{black}{At an assumed HVS ejection rate of $10^{-4} \, \mathrm{yr^{-1}}$, HVSs detectable in the final \textit{Gaia} data release and LSST from the LMC (MW) will number $125_{-12}^{+11}$ ($50_{-8}^{+7}$) and $140_{-11}^{+10}$ ($42_{-7}^{+6}$), respectively.} The MW and LMC HVS populations show different kinematics and spatial distributions. While LMC HVSs have more modest total velocities and larger Galactocentric distances clustered around those of the LMC itself, HVSs from the MW show broader distributions, including a prominent high-velocity tail above $500 \, \mathrm{km \ s^{-1}}$ that contains at least half of the stars. These predictions are robust against reasonable variation of the Galactic potential and of the LMC central black hole mass.
\end{abstract}




\section{Introduction}
Hyper-velocity stars (HVSs) are stars moving so fast that they are gravitationally unbound from the Galaxy. The existence of such a population of stars was first proposed by \citet{Hills1988}, who theorized that a three-body interaction between a binary system and a massive black hole (MBH) at the centre of the Milky Way could disrupt the binary and eject one companion at $\sim 1000 \, \mathrm{km \ s^{-1}}$. The serendipitous discovery by \citet{Brown2005} of a $3 \, \mathrm{M_\odot}$ main sequence star in the outer Galactic halo unbound to the Galaxy by its radial velocity alone ($v_{\rm rad} \simeq 830 \, \mathrm{km \ s^{-1}}$, see \citealt{Brown2014}) was touted as the first direct observational evidence in favour of this `Hills mechanism' and of hyper-velocity stars in general. Further detections of unbound HVS candidates followed soon after \citep{Hirsch2005, Edelmann2005}. To date, several dozen HVS candidates have been discovered via targeted searches in the Galactic halo \citep[e.g.][]{Brown2006, Brown2009, Brown2012, Brown2014} or via searches in data releases of ongoing large surveys \citep[e.g.][]{Zhong2014, Huang2017, Li2018, Hattori2018, Marchetti2019, Koposov2020, Marchetti2021}. See \citet{Brown2015rev} for a review of these objects. 

HVSs are intriguing tools to study a variety of astrophysical phenomena -- the violent and uncertain physical processes that generate them leave an imprint on their kinematics and properties. As objects born in the GC but located in more observationally accessible regions of the sky, HVSs offer a `back door' into investigating the heavily dust-obscured and source-crowded GC environment \citep[see][]{Zhang2013, Madigan2014, Rossi2017}, e.g. the nature of the nuclear star cluster \citep[see][for a review]{Boker2010} and the growth of Sgr A* and its impact on its environment \citep{Genzel2010, Bromley2012}. The deceleration and deflection of HVSs during their flights from the GC to the outer halo make them promising dynamical tracers, providing constraints on the shape, size and mass of the Galactic dark matter halo \citep[e.g.,][]{Gnedin2005, Yu2007, Kenyon2008, Kenyon2014, Contigiani2019}.

It is worth noting that for most candidate HVSs, existing astrometry is not precise enough to indisputably associate them with an origin in the Galactic Centre (GC)\footnotemark. However, especially given the high-precision astrometry provided by the European Space Agency's \textit{Gaia} mission \citep{Gaia2016,Gaia2018,Gaia2020EDR3}, the GC can be confidently \textit{excluded} as the birthplace of many HVS candidates \citep[see e.g.][]{Irrgang2018, Kreuzer2020}. The past trajectories of some candidates seem to point towards the Galactic disc \citep[e.g.,][]{Heber2008,Silva2011, Palladino2014, Irrgang2018, Irrgang2019, Marchetti2018}. For these disc-ejected main sequence `hyper-runaway' stars, it is currently unclear which physical mechanism(s) are responsible for ejecting them \citep[see discussion in][]{Evans2020}. Often-blamed mechanisms include dynamical encounters in dense systems \citep{Poveda1967, Leonard1990, Leonard1991, Perets2012, Oh2016} and the disruption of a tight binary following a core-collapse event \citep{Blaauw1961,Tauris1998, Portegies2000, Tauris2015, Renzo2019, Evans2020}.

\footnotetext{The notable exception to this is the HVS candidate S5-HVS1 \citep{Koposov2020}, whose short flight time and precise astrometry allow it to be traced back definitively to the GC.}

The kinematics and/or ages of other HVS candidates do not suggest an origin in the inner Milky Way at all. They either have past trajectories pointing away from the Galaxy entirely \citep[see][]{Marchetti2019} and/or have stellar lifetimes too short to accommodate a journey from the inner Milky Way to their current position. The latter is the case for HVS3 \citep{Edelmann2005}, an $8 \, \mathrm{M_\odot}$ HVS candidate at a distance of $61 \, \mathrm{kpc}$. Using \textit{Gaia} Data Release 2 astrometry, \citet{Erkal2019} found its flight time to Galactic pericentre to be $\sim 66 \, \mathrm{Myr}$, in tension with its nominal main sequence lifetime of only $\sim 35 \, \mathrm{Myr}$. It was noted as far back as \citet{Edelmann2005}, however, that HVS3 is situated only $16.3^{\circ}$ from the Large Magellanic Cloud (LMC). An LMC origin for HVS3 was further suggested by \citet{Gualandris2007}, who simulate ejections via the Hills mechanism of HVS3-like stars from the LMC, and by \citet{Przybilla2008HVS3}, who find elemental abundances in HVS3 more similar to reference stars in the LMC than to reference stars in the GC. Indeed, \citet{Erkal2019} find that the past trajectory of HVS3 tracks directly to the LMC centre and requires a flight time of only $\sim 21 \, \mathrm{Myr}$. This, along with its putative ejection velocity of $\sim 870 \, \mathrm{km \ s^{-1}}$ relative to the LMC centre at its point of closest approach, strongly support a Hills mechanism ejection for HVS3, requiring an LMC MBH mass of at least $10^{3.6} - 10^{4} \, \mathrm{M_\odot}$.

With the confirmation of HVS3 as an HVS of Magellanic origin, it is natural to wonder how many other LMC-ejected HVSs are out there. \textcolor{black}{In light of the ever-increasing population of HVS candidates, it is important to characterize the contribution of the LMC to this population. While the rate at which HVSs are ejected from the LMC centre is currently unconstrained, the LMC remains arguably the most promising extra-Galactic source of HVSs. It is the nearest galaxy massive enough to host an MBH, and its large orbital velocity in the Galactocentric rest frame ($\sim320 \, \mathrm{km \ s^{-1}}$; \citealt{Kallivayalil2013}) and modest escape velocity ($\sim 100 \, \mathrm{km \ s^{-1}}$ from the centre  to $2 \, \mathrm{kpc}$ in our default MW+LMC potential) mean that its ejected stars can quite easily escape the inner LMC. Other possible extra-Galactic provenances of HVSs include HVSs ejected from M31 or its satellites \citep{Lu2007,Sherwin2008}, or HVSs tidally stripped from infalling dwarf galaxies \citep{Abadi2009}.}

To date, HVS3 remains the only HVS conclusively associated with the LMC. This may change, however, with the advent of massive Galactic surveys targeting billions of Milky Way sources such as the \textit{Gaia} mission and the Vera C. Rubin Observatory's Legacy Survey of Space and Time \citep[LSST; see][]{Ivezic2019}. In this work we realistically simulate the ejection of HVSs from MBHs located in both the LMC centre and the GC. We propagate these ejected HVSs through a Galactic potential consisting of the Milky Way and a moving LMC. \textcolor{black}{Making reasonable assumptions about the \textit{Gaia} and LSST selection functions, we predict the size and properties of the GC HVS and LMC HVS populations in future data releases from these surveys}. By identifying contrasts between the two populations, we provide insight on how LMC HVSs can be most efficiently unearthed. We additionally explore how these predictions depend on the MW+LMC potential and the LMC MBH mass. 

\textcolor{black}{This work follows and complements earlier research into the ejection of HVSs from the LMC. Our method of ejecting and propagating stars from the LMC resembles broadly \citet{Boubert2016}, who eject $3 \, \mathrm{M_\odot}$ HVSs via the Hills mechanism from the LMC centre to explore their sky distribution on the sky, agnostic of the HVS ejection rate. They predict a dipolar distribution of LMC HVSs -- a cluster near the present day LMC location and a corresponding dearth of LMC HVSs in the Galactic northeast. Extending their work, we employ a more comprehensive ejection model and exploit mock \textit{Gaia} and LSST observations to make quantitative predictions on the number of detectable LMC HVSs under an assumed ejection rate. Exploring the possibility that HVSs are ejected from the LMC disc following binary supernovae, \citet{Boubert2017} find that this mechanism can contribute to the known population of Galactic HVS candidates at a rate of $\sim3\times10^{-6} \, \mathrm{yr^{-1}}$, many of which will be observable by \textit{Gaia} (see also Sec. \ref{sec:disc:runaways}).}

In Sec. \ref{sec:methods} we describe our HVS ejection model, our parametrization of MW+LMC potential, our orbital integration, our approach to obtaining mock photometric observations of our sample, and the \textit{Gaia} and LSST selection cuts we employ. In Secs. \ref{sec:results:allstars}, \ref{sec:results:cuts} and \ref{sec:potentialresults} we present our results and discuss these results in Sec. \ref{sec:discussion}. We present our conclusions in Sec. \ref{sec:conclusions}. To clarify the terminology used in this work, we use for simplicity the term HVS to refer to \textit{any and all} stars ejected from the GC or LMC centre via interactions with an MBH, regardless of ejection velocity. This is a more general use of the term than the convention, which typically refers exclusively to stars unbound to the Milky Way ejected from the GC \citep[e.g.][]{Brown2015rev}. 

\section{Mock Ejected Star Catalogues} \label{sec:methods}

The generation of realistic mock populations of GC and LMC HVSs requires a number of ingredients. In the following subsections we outline how we draw stellar velocities and masses at the moment of ejection, how we assign ages and flight times to ejected stars, how we propagate ejected stars forward in time, and how we obtain mock photometric and astrometric observations for these populations. If not otherwise stated, the methodology is the same for both GC and LMC HVSs.

\subsection{Ejection distribution and rate}
\label{sec:methods:ejection}
We generate populations of Hills mechanism-ejected stars from GC and the LMC centre following a Monte Carlo ejection model similar to that outlined in \citet{Rossi2017}, \citet{Marchetti2017} and implemented in \citet{Marchetti2018}. We describe the model as follows.

We generate binary systems defined by the mass of the primary $m_{\rm p}$, the mass ratio between the secondary and primary $q$ and the orbital semi-major axis $a$. Following an interaction with the MBH, the binary is disrupted -- one member star is ejected while the other remains bound to the MBH. \citet{Sari2010} and \citet{Kobayashi2012} show that for a parabolic approach, there is an equal probability for ejecting either star in the binary. We therefore randomly designate one star among the pair as the ejected star. Its ejection velocity is calculated analytically \citep{Sari2010, Kobayashi2012,Rossi2014}:
\begin{equation}
    v_{\rm ej} = \sqrt{\frac{2Gm_{\rm c}}{a}} \left( \frac{M_{\rm BH}}{M} \right)^{1/6} \, \text{,}
    \label{eq:vej}
\end{equation}
where $M=(1+q)m_{\rm p}$ is the total mass of the projenitor binary, $m_{\rm c}$ is the mass of the remaining bound companion, and $M_{\rm BH}$ is the MBH mass. For stars ejected from the GC, we take $M_{\rm BH, GC}=4\times10^{6} \, \mathrm{M_\odot}$ \citep{Eisenhauer2005, Ghez2008}. For ejections from the LMC centre, we assume in our fiducial model an LMC MBH mass of $10^{5} \, \mathrm{M_\odot}$, similar to the median MBH mass quoted by \citet{Reines2013}, who study dwarf galaxies with stellar masses similar to the LMC ($10^{8.5}\lesssim M_{*}\lesssim 10^{9.5} \, \mathrm{M_\odot}$). We explore in Sec. \ref{sec:discussion} the sensitivity of our results to variation in $M_{\rm BH, LMC}$.

We draw primary masses for the binaries assuming a \citet{Kroupa2001} mass function in the range [0.1,100] $\mathrm{M_\odot}$, consistent with observations of the central regions of the Milky Way \citep{Bartko2010,Lockmann2010} and broadly consistent with observations of LMC clusters at least in the $m>0.5 \, \mathrm{M_\odot}$ regime \citep{DaRio2009,Liu2009NGC1818,Liu2009}. Mass ratios are sampled in the interval $0.1\leq q \leq 1 $ distributed as $f(q)\propto q^{-3.5}$ and semi-major axes follow \"{O}pik's law, $f(a)\propto a^{-1}$ \citep{Opik1924}, both consistent with \citet{Dunstall2015} who investigate B-type binaries in the VLT-FLAMES Tarantula Survey of the 30 Doradus region in the LMC \citep{Evans2011}. Modest variations in the power law indices for the distributions described above, including switching to a more top-heavy GC initial mass function (IMF) as suggested by \citet{Lu2013}, do not significantly impact the results of this study. Binary orbital semi-major axes are generated in the range $2.5\, \text{max}(R_{\rm p}, R_{\rm s})\leq a \leq 2000 \, \mathrm{R_\odot}$, where $R_{\rm p}$ and $R_{\rm s}$ are the stellar radii of the primary and secondary stars, respectively. This lower limit is set by Roche lobe overflow. Finally, we remove stars less massive than $0.5 \, \mathrm{M_\odot}$, as these ejected stars will more than likely be too faint for current or near-future surveys to detect.

\begin{figure}
    \centering
    \includegraphics[width =\columnwidth]{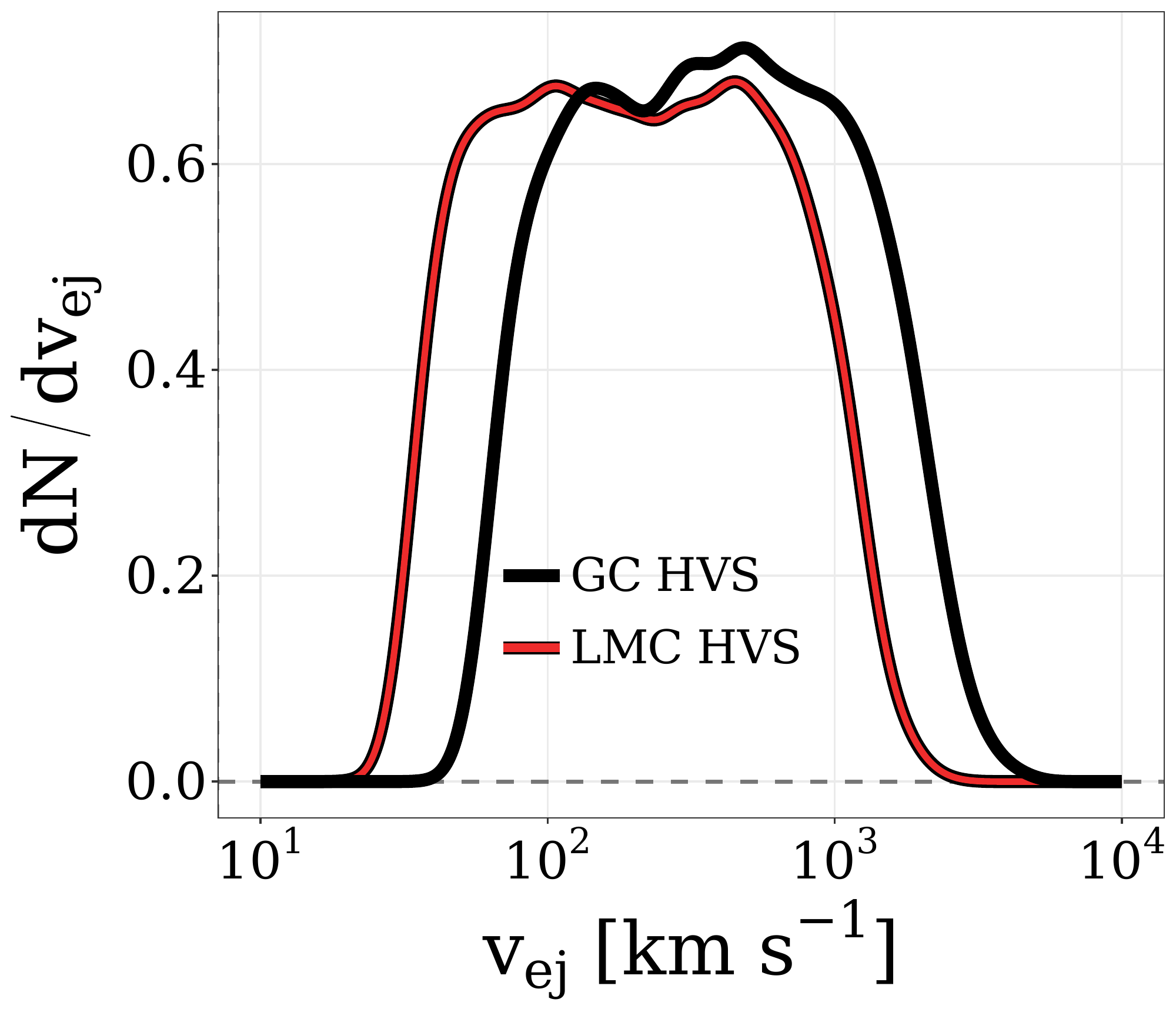}
    \caption{Distribution of stellar ejection velocities via the Hills mechanism (see Sec. \ref{sec:methods:ejection}) in the rest frame of the MBH in the LMC centre (red; $M_{\rm BH, LMC}=1\times10^5 \, \mathrm{M_\odot}$) and GC (black; $M_{\rm BH, GC}=4\times10^{6} \, \mathrm{M_\odot}$).}
    \label{fig:vejs}
\end{figure}

We show in Fig. \ref{fig:vejs} the distributions of ejection velocities resulting from Eq. \ref{eq:vej} and the above assumptions for HVS ejections from both the GC and LMC. The median HVS ejection velocity from the GC is $370 \, \mathrm{km \ s^{-1}}$. Note that in our MW+LMC potential (see Sec. \ref{sec:methods:potential}), the escape velocity from the GC to $2 \, \mathrm{kpc}$ is $\sim 600 \, \mathrm{km \ s^{-1}}$ -- only 36 per cent of stars are ejected with sufficient velocity to escape the Galactic bulge. The median ejection velocity from the LMC centre is $200 \, \mathrm{km \ s^{-1}}$ in the LMC rest frame. Despite lower typical ejection velocities, $\sim$70 per cent of stars ejected from the LMC centre will escape the inner LMC owing to its modest effective escape velocity ($100 \, \mathrm{km \ s^{-1}}$ to $2 \, \mathrm{kpc}$ in our MW+LMC potential). 

Generated HVSs are ejected isotropically on initially radial trajectories in the Galactocentric and LMC-centric rest frames for the GC and LMC HVS populations, respectively. For LMC ejections we assume the LMC MBH is coincident with our assumed LMC centre  at $\alpha = 78.76^{\circ}, \, \delta = -69.19^{\circ}$ \citep{Zivick2019}. In at least a substantial fraction of dwarf galaxies, the MBH can be offset by the host dwarf centre by as much as a few kiloparsecs \citep[see][]{Reines2020, Mezcua2020, Bellovary2021}. A modest deviation of the LMC MBH in position and velocity from the assumed LMC centre does not appreciably affect our predictions for the number of GC and LMC HVSs lurking in future surveys. 

Theoretical estimates and observational constraints imply an HVS ejection rate from the GC of $\sim10^{-5} - 10^{-3} \, \mathrm{yr^{-1}}$ \citep[see][]{Brown2015rev}. We adopt a constant ejection rate of $10^{-4} \, \mathrm{yr^{-1}}$ from the GC in all models. \textcolor{black}{In light of the lack of observational constraints on the putative LMC MBH and its stellar environment, we assume the HVS ejection rate from the LMC centre to be $10^{-4} \, \mathrm{yr^{-1}}$ as well. We discuss further the uncertainty surrounding the HVS ejection rate in Sec. \ref{sec:disc:rates}}. Our characterization of the spatial distribution and kinematics of HVS candidates observable by these surveys will be stacked and averaged over multiple ejection samples, and will thus not depend on the ejection rate. 

\subsection{HVS flight time distribution} \label{sec:methods:tflight}

In addition to masses and initial velocities, we assign each ejected star a flight time and a stellar age at ejection. For HVS ejections from the GC we consider ejections throughout the entire lifetime of the Milky Way, assumed here to have formed shortly after the Big Bang 13.8 Gyr ago \citep{Planck2016}. The LMC is most likely on its first infall, having first crossed the virial radius some $\sim1.5-4 \, \mathrm{Gyr}$ ago \citep{Besla2007, Kallivayalil2013, Gomez2015, Patel2017}. We therefore consider ejections from the LMC over the last 4 Gyr\footnote[2]{The precise lookback time up to which we consider LMC HVS ejections is not particularly important when predicting HVS populations in \textit{Gaia} or LSST -- stars with flight times $>1 \, \mathrm{Gyr}$ are generally too far (and thus too dim) to be detectable by these surveys (see Sec. \ref{sec:results:cuts})}. 

Our present day mock catalogue of HVSs is composed of stars ejected $t_{\rm ej}$ ago that have not yet left the main sequence. We assign $t_{\rm ej}$ uniformly: 
\begin{align}
    t_{\rm ej} &= \epsilon_1 \cdot \mathrm{4 \, Gyr} &\; &\text{from LMC}\\
    t_{\rm ej} &= \epsilon_1 \cdot \mathrm{13.8 \, Gyr} &\; &\text{from GC,}
\label{eq:tflight}
\end{align}
where $0<\epsilon_1 < 1$ is a uniform random number.
We assume no preferred stellar age at time of ejection. The age of a star at ejection $t_{\rm age, ej}$ is then a random fraction $\epsilon_2$ of its main sequence lifetime $t_{\rm MS}$;
\begin{equation}
    t_{\rm age, ej} = \epsilon_2 \cdot t_{\rm MS} \; ,
\end{equation}
where $t_{\rm MS}$ is calculated according to the stellar mass and assumed metallicity using the \citet{Hurley2000} analytical formulae. Though stars in the GC show a large spread in metallicity \citep{Do2015, FeldmeierKrause2017, Rich2017}, for GC HVSs we assume solar metallicity, $Z_{\odot}=0.02$ \citep{Anders1989}. For the LMC HVSs we assume a lower metallicity of $\mathrm{log_{10}}[Z/Z_\odot] = -0.5$ \citep[see][and references therein]{Moe2013,Piatti2013, Choudhury2016}. 

The remaining main sequence lifetime of the star $t_{\rm left}$ is
\begin{equation}
    t_{\rm left} = t_{\rm MS} - t_{\rm age, ej} = (1-\epsilon_2) \cdot t_{\rm MS} \; .
\end{equation}
We restrict our analysis to main sequence stars to more easily compare with existing HVS candidates. We therefore remove stars for whom $t_{\rm ej}>t_{\rm left}$. The flight time of each mock HVS is then
\begin{equation}
    t_{\rm flight} = t_{\rm ej}
\end{equation}
and its current age is
\begin{equation} \label{eq:tage}
    t_{\rm age,0} = t_{\rm age,ej} + t_{\rm flight} \; .
\end{equation}
These stellar ages are useful in obtaining mock photometry (see Sec. \ref{sec:methods:observations}).

After assigning masses, ejection velocities, flight times and ages to each HVS according to Eqs. \ref{eq:vej}-\ref{eq:tage}, removing $m<0.5 \, \mathrm{M_\odot}$ stars and stars which don't survive on the main sequence until the present day, we are left with samples of $\sim$70,000 GC HVSs and $\sim$37,000 LMC HVSs. We draw, eject and propagate 50 HVS samples from each origin to eliminate stochastic effects on our predictions. 

\subsection{Milky Way + LMC potential} \label{sec:methods:potential}

We describe here the MW + moving LMC potential through which we propagate our ejected HVS. We explore variations in these assumptions in Sec. \ref{sec:discussion}. We model the MBH in the Galactic Centre by a Keplerian potential, 
\begin{equation}
    \label{kepler}
    \Phi_{\rm BH, GC}(r) = -G \, \frac{4\times10^{6} \, \mathrm{M_\odot}}{r} \; ,
\end{equation}
where $r$ is the Galactocentric distance. The Galactic bulge is treated as a Hernquist potential \citep{Hernquist1990},
\begin{equation}
    \label{hernquist}
    \Phi(r) = -\frac{G M_{\rm b} }{r+r_{\rm b}} \; ,
\end{equation}
where $M_{\rm b}=3.4 \times 10^{10} \, \mathrm{M_{\odot}}$ and $r_{\rm b}=0.7 \, \mathrm{kpc}$ are the bulge scale mass and radius, respectively \citep{PriceWhelan2014}. For the Galactic disc we use a Miyamoto-Nagai potential in Galactocentric cylindrical coordinates, \citep{Miyamoto1975},
\begin{equation}
    \Phi(R,z) = -\frac{G M_{\rm d}}{\sqrt{R^2+(a_{\rm d}+\sqrt{z^2+b_{\rm d}^2}) }} \; ,
\end{equation}
where $M_{\rm d}$ is the disc mass and $a_{\rm d}$ and $b_{\rm d}$ are the disc scale length and height, respectively. We use $M_{\rm d} = 10^{11} \, \mathrm{M_{\odot}}$, $a_{\rm d} = 6.5 \, \mathrm{kpc}$ and $b_{\rm d} = 260 \, \mathrm{pc}$ following \citet{PriceWhelan2014}.

We model the contribution from Milky Way dark matter halo using a spheroidal NFW potential \citep{Navarro1996, Navarro1997}:
\begin{equation}
    \rho(x,y,z) = \frac{M_{\rm s}}{4\pi r_{\rm s}} \frac{1}{(\xi/r_{\rm s})(1+\xi/r_{\rm s})^2} \, , \: \;  \xi^2 = x^2\text{+}y^2\text{+}\frac{z^2}{c_{\rm halo}} \, ,
\end{equation}
where $M_{\rm s}$ and $r_{\rm s}$ are the scale mass and radius of the halo, respectively, and $c_{\rm halo}$ is the dimensionless z-to-x axis ratio of the spheroid. In our fiducial potential we assume $c_{\rm halo}=1$ and choose $M_{\rm s} = 0.76\times 10^{12} \, \mathrm{M_{\odot}}$ and $r_{\rm s} = 24.8 \, \mathrm{kpc}$ -- the best fit parameters to the Galactic rotation curve for a spherical halo from \citet{Rossi2017}. 

We model the LMC potential as a two-component Keplerian MBH + Hernquist bulge. We assume an LMC MBH mass of $10^{5} \, \mathrm{M_\odot}$ (see Sec. \ref{sec:methods:ejection}). As a fiducial LMC scale mass we choose $M_{\rm LMC}=1.5 \times 10^{11} \, \mathrm{M_\odot}$ and a corresponding scale radius of $R_{\rm LMC} = 17.14 \, \mathrm{kpc}$. This is a common choice for $M_{\rm LMC}$ \citep[e.g.][]{Erkal2019, Belokurov2019, Erkal2020}, consistent with recent determinations \citep{Kallivayalil2013, Penarrubia2016, Laporte2018, Erkal2019LMCmass, Erkal2020LMCmass}. In each simulation, the LMC is initialized at its observed position and velocity today. We use a centre-of-mass position of $\alpha = 78.76^{\circ}, \delta = -69.19^{\circ}$ \citep{Zivick2019}, proper motions of ($\mu_{\alpha^*},\mu_\delta) = (1.91, 0.229) \, \mathrm{mas \ yr^{-1}}$ \citep{Kallivayalil2013}, heliocentric distance of $49.97 \, \mathrm{kpc}$ \citep{Pietrzynski2013} and radial velocity of $+262.2 \, \mathrm{km \ s^{-1}}$ \citep{vanderMarel2002}.

\subsection{Orbital Integration} \label{sec:methods:integration}

In both GC and LMC centre ejections, stars are initialized at a random point on the surface of the MBH's so-called `sphere of influence', where the contribution of the MBH to the total potential becomes subdominant. For the GC MBH we adopt a sphere of influence radius of $3 \, \mathrm{pc}$ \citep{Genzel2010}. In the absence of constraints on the size of the sphere of influence of the LMC MBH, we adopt this same value. The results of this study are not sensitive to modest variations of this radius.

For each GC HVS and LMC HVS, the LMC is first integrated backwards in time to the star's flight time $t_{\rm flight}$ ago.  Stars ejected from the GC are initialized with velocities pointing directly away from the GC. Similarly, LMC-ejected stars are initially placed $3 \, \mathrm{pc}$ from the LMC centre at the appropriate time and their initial velocities in the Galactocentric rest frame are the vector addition of their ejection velocities in the LMC rest frame and the LMC orbital velocity $t_{\rm flight}$ ago. The LMC and the HVS are both propagated forward in time until the present day. Orbits are integrated in the \texttt{PYTHON} package \texttt{GALPY}\footnote[3]{\url{https://github.com/jobovy/galpy}} \citep{Bovy2015} using a fifth-order Dormand-Prince integrator \citep{Dormand1980} with a fixed timestep of $0.1 \, \mathrm{Myr}$. We assume the Sun is $8.178 \, \mathrm{kpc}$ from the GC \citep{Gravity2019} and $25 \, \mathrm{pc}$ above the Galactic disc \citep{Bland-Hawthorn2016}. We assume a circular velocity at the Solar position of $232.76 \, \mathrm{km \ s^{-1}}$ \citep{McMillan2017} and take $(U_{\odot}, V_{\odot}, W_{\odot})=(14.0, 12.24, 7.25) \, \mathrm{km \ s^{-1}}$ \citep{Schonrich2012}. The impact of dynamical friction on the LMC's orbit is accounted for using the implementation in \texttt{GALPY} which follows roughly \citet{Petts2016}.

\subsection{Mock photometric \& astrometric observations} \label{sec:methods:observations}

\begin{figure*}
    \centering
    \includegraphics[width =1.4\columnwidth]{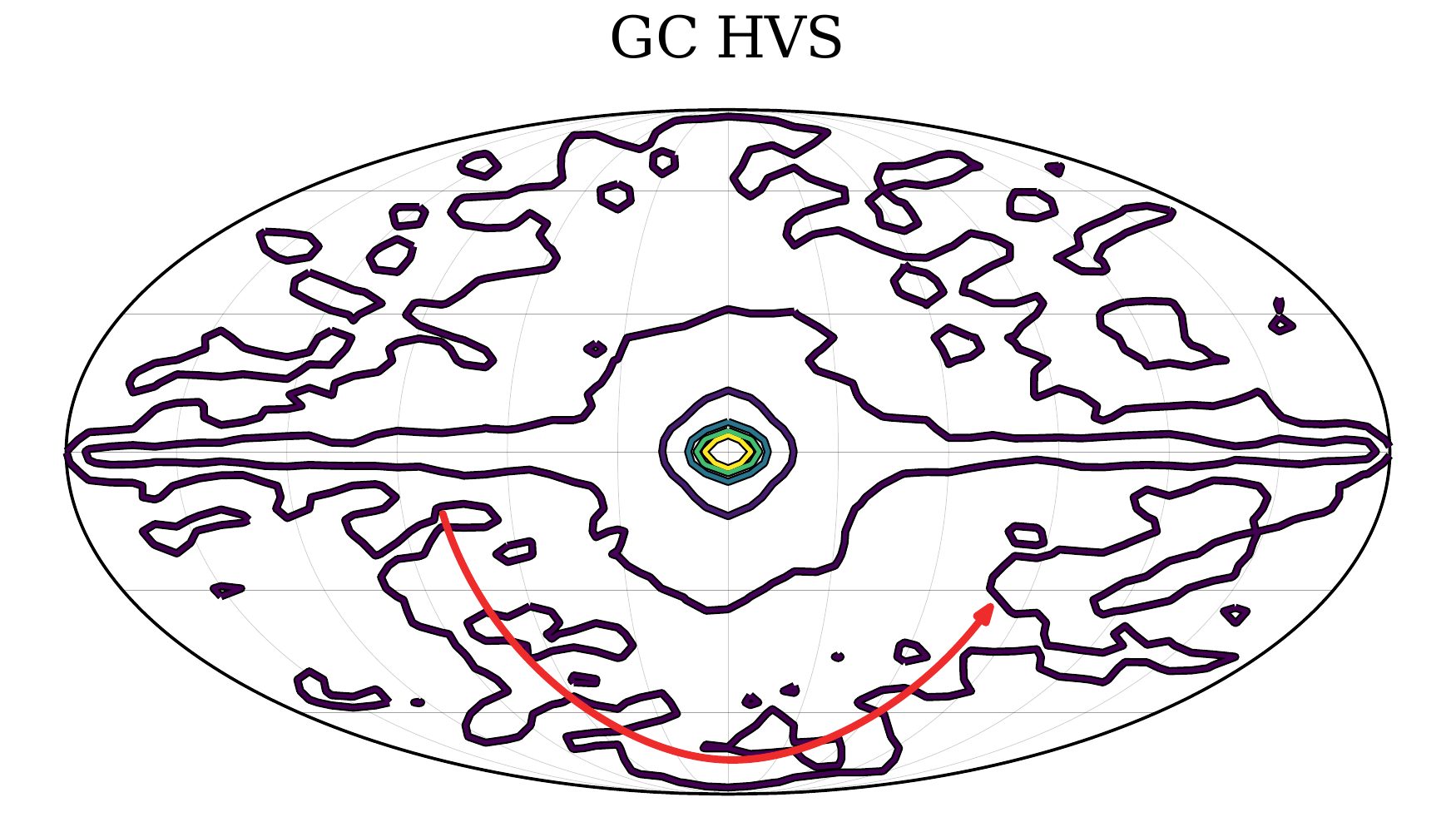}
    \includegraphics[width =1.4\columnwidth]{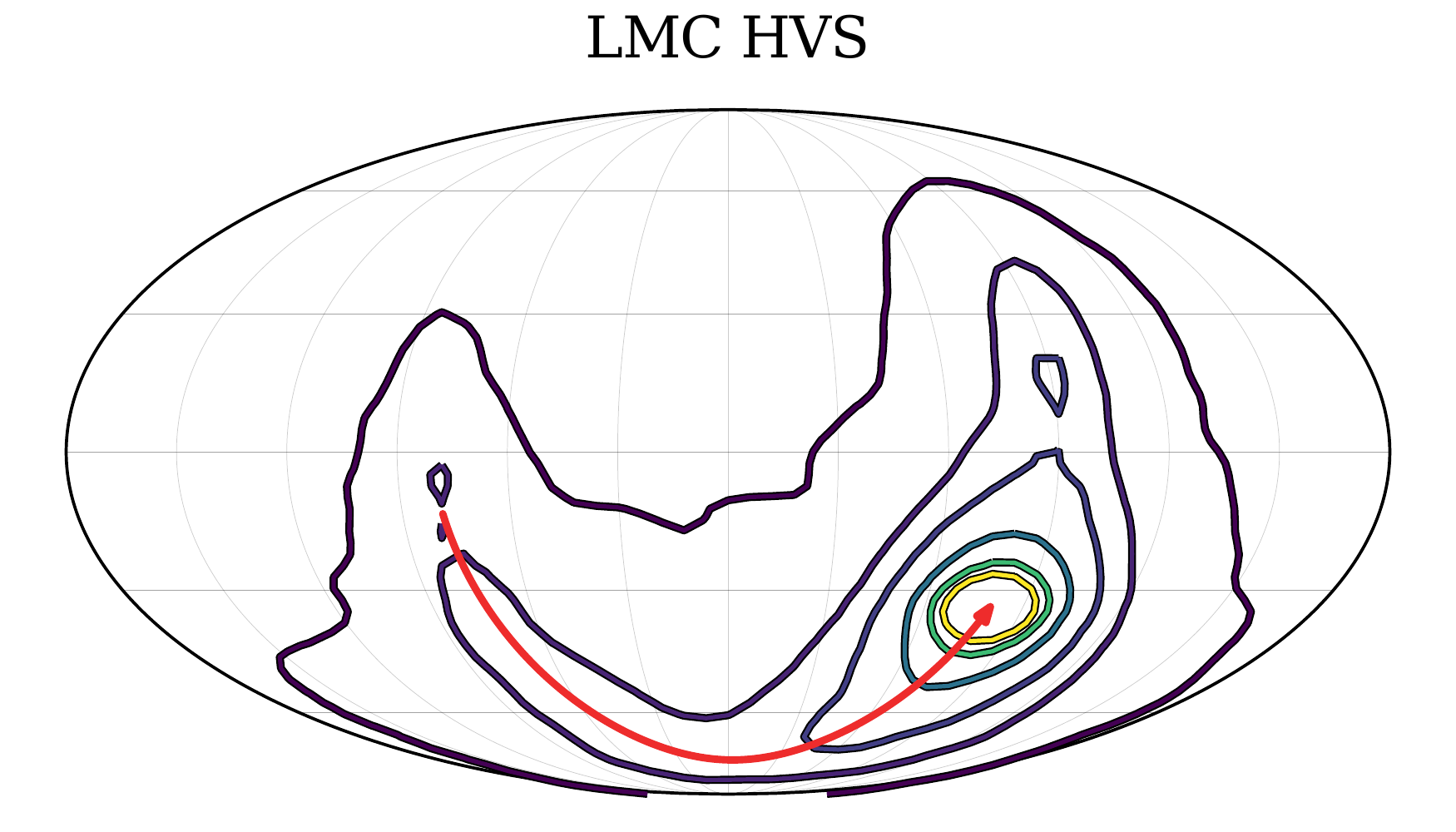}
    \caption{Density contours in Galactic coordinates using a Mollweide projection for HVSs ejected via the Hills mechanism from the Galactic Centre (left) and LMC centre (right). Contours are spaced logarithmically. The red curve shows the LMC trajectory on the sky over the last Gyr.}
    \label{fig:skycontour}
\end{figure*}

After propagating our populations of mock HVSs, we compute photometric properties following \citet{Marchetti2018}. Given the mass, metallicity and age of the star today, we obtain its physical radius, effective temperature and surface gravity using the analytic formulae of \citet{Hurley2000}. Via chi-squared minimization of the effective temperature and surface gravity, we find the best-fitting stellar spectrum among the $\mathrm{log_{10}}[Z/Z_\odot] =0.0$ spectra (for GC HVSs) or the $\mathrm{log_{10}}[Z/Z_\odot]=-0.5$ spectra (for LMC HVSs) in the BaSeL SED Library 3.1 \citep{Westera2002}. We assume an atmospheric micro-turbulence velocity of $2 \ \mathrm{km \, s^{-1}}$ and a mixing length of zero. We calculate the visual extinction $A_{\rm V}$ at each star's position using the \texttt{MWDUST}\footnote[4]{\url{https://github.com/jobovy/mwdust}} three-dimensional Galactic dust map \citep{Bovy2016}; itself a combination of the \citet{Drimmel2003}, \citet{Marshall2006}, and \citet{Green2015} maps. The attenuation at all wavelengths $A_{\lambda}$ is calculated from the visual extinction assuming a \citet{Cardelli1989} reddening law with $R_{\rm V}=3.1$. We compute each star's magnitude in the \textit{Gaia G} filter by integrating the flux of the best-fitting BaSeL spectrum and the attenuation $A_{\lambda}$ through the \textit{Gaia} Early Data Release 3 (EDR3) $G$ passband\footnote[5]{\url{https://www.cosmos.esa.int/web/gaia/edr3-passbands}} \citep[see][eq. 1]{Jordi2010}. We similarly compute magnitudes in the Johnson-Cousins \textit{I$_{C}$} and \textit{V} bands, adopting the \citet{Bessell1990} passbands curves. From these, the magnitude in the \textit{Gaia G$_{\rm RVS}$} band can be computed using the polynomial fits in \citet{Jordi2010}. We similarly compute magnitudes in the LSST $r$ band using the expected throughput of the LSST $r$ passband\footnote[6]{\url{https://github.com/lsst/throughputs}}. We use the \texttt{PYTHON} package \texttt{PyGaia}\footnote[7]{\url{https://github.com/agabrown/PyGaia}} to estimate the end-of-mission \textit{Gaia} parallax and proper motion uncertainties for each HVS. 

\section{HVS space and velocity distributions} \label{sec:results:allstars}

\begin{figure*}
    \centering
    \includegraphics[width =1\columnwidth]{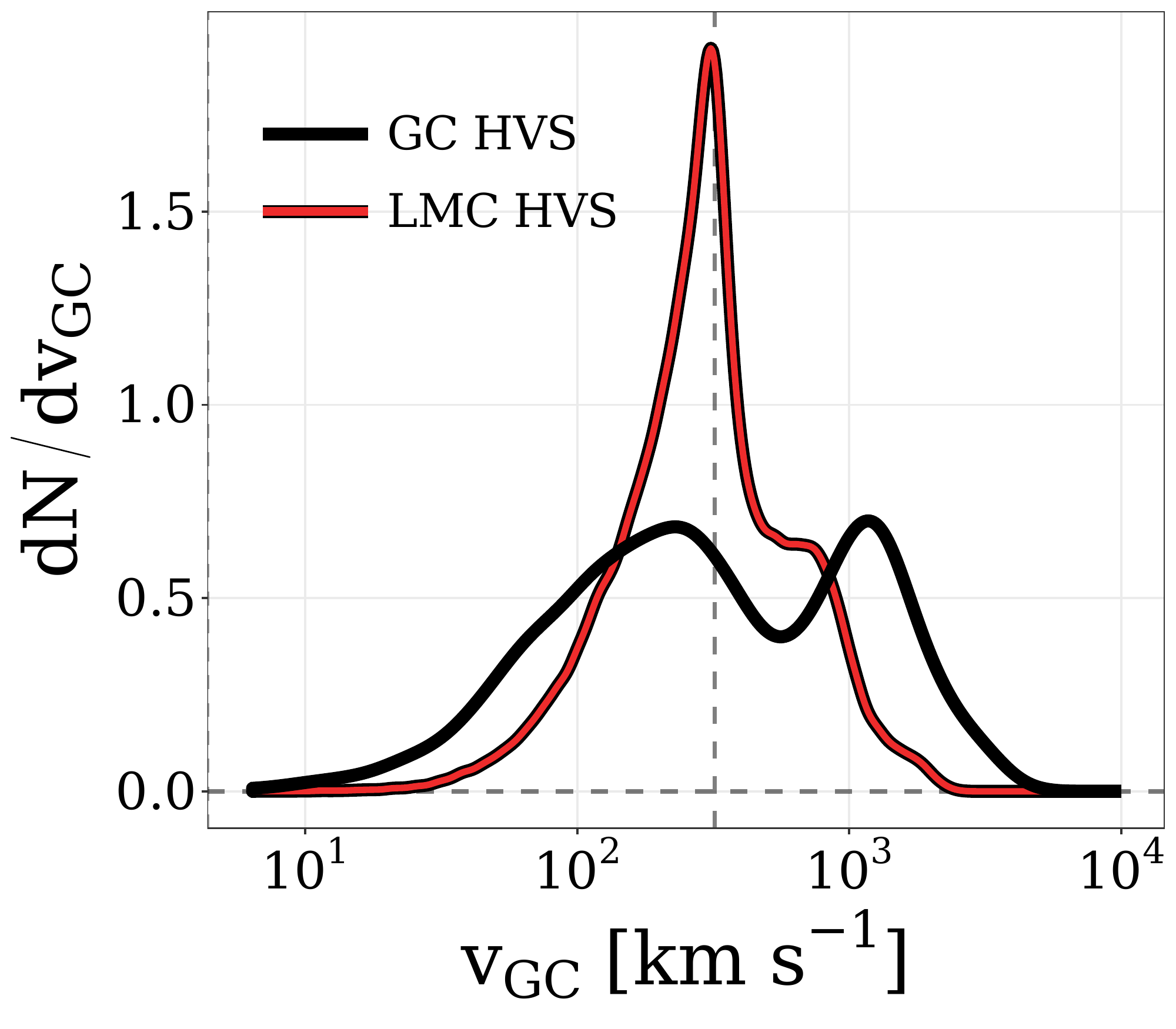}
    \includegraphics[width =1\columnwidth]{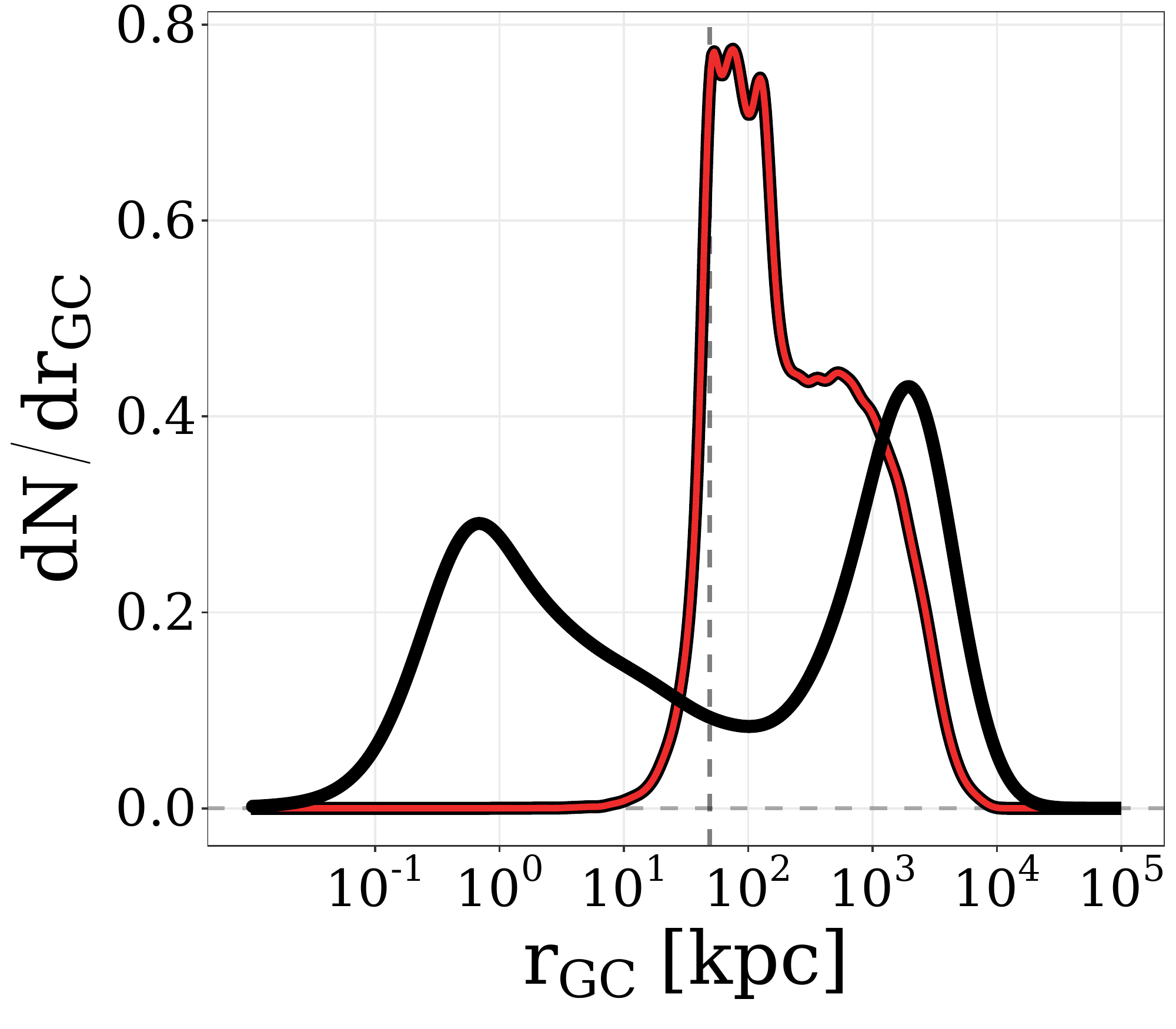}
    \caption{Distributions of Galactocentric velocities (left) and distances (right) for $m>0.5 \, \mathrm{M_\odot}$ HVSs ejected from the GC (black) and LMC (red) propagated through the MW+LMC potential. Vertical dashed lines show the Galactocentric velocity ($320 \, \mathrm{km \ s^{-1}})$ and distance ($49 \, \mathrm{kpc}$) of the LMC today.}
    \label{fig:GCv_GCdist}
\end{figure*}

\begin{figure*}
    \centering
    \includegraphics[width =2\columnwidth]{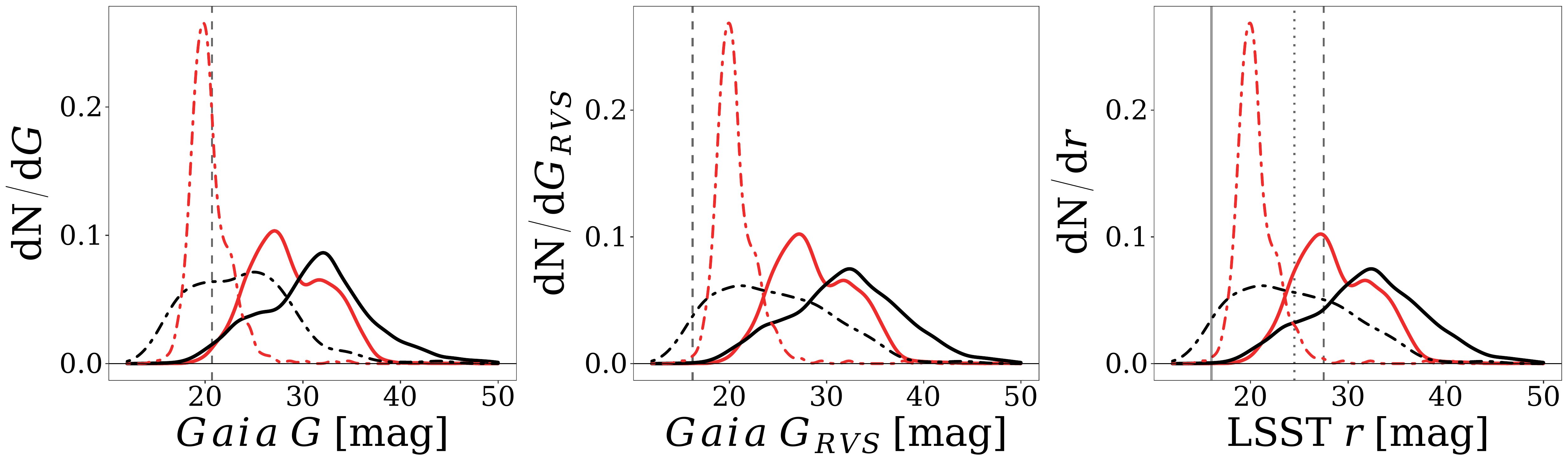}
    \caption{Apparent magnitude distributions for HVSs in the \textit{Gaia G} (left), \textit{Gaia G$_{\rm RVS}$} (center) and LSST \textit{r} (right) bands. Red and black curves show LMC HVS and GC HVS populations, respectively. Solid and dash-dotted curves show distributions for $m>0.5 \, \mathrm{M_\odot}$ and $m>2 \, \mathrm{M_\odot}$ HVSs, respectively. Dashed vertical lines in the left and center panels correspond to $G=20.7$ and $G_{\rm RVS}=16.2$; the magnitude limits for the \textit{Gaia} final data release five-parameter astrometric catalogue and the radial velocity catalogue, respectively. Solid, dotted and dashed vertical lines in the right panel correspond to $r=16$, $r=24.5$ and $r=27.5$; the estimated LSST saturation limit, single-visit magnitude limit and ten-year coadded magnitude limit, respectively.} 
    \label{fig:magnitudes}
\end{figure*}

In this Section we explore the space and velocity distributions of all HVSs ejected from the GC and LMC, regardless of \textcolor{black}{they are detectable by \textit{Gaia} or LSST.}
 
In Fig. \ref{fig:skycontour} we show the spatial distribution of GC HVSs and LMC HVSs on the sky. While no region of the sky is entirely inaccessible by LMC HVSs (lower panel), the vast majority of stars are in close proximity to the LMC. A notable feature is the arm of HVSs that leads the LMC orbit. The origin of this arm is the fact that stars ejected from the LMC along its direction of motion will be boosted by its $v_{\rm GC}\simeq320 \, \mathrm{km \ s^{-1}}$ orbital velocity in the MW rest frame \citep{Kallivayalil2013}, reaching higher Galactocentric velocities than stars ejected on angles misaligned with the LMC's orbital motion. These cluster and arm features are consistent with the findings of \citet{Boubert2016}, who suggest the leading arm of LMC HVSs as a possible explanation for the apparent excess of known HVS candidates in the vicinity of Leo \citep[see][]{Brown2009Leo}. There is also a tail of HVSs lagging the LMC orbit on the sky, though we caution this tail consists mainly of very distant, long-$t_{\rm flight}$ stars which will not be observable by \textit{Gaia} or LSST (see following Section).

Fig. \ref{fig:skycontour} suggests that the vast majority of HVSs ejected from the GC (upper panel) will be in close proximity to the GC on the sky and otherwise mostly confined to the Galactic midplane. Due to the large escape velocity from the inner Galaxy, only the HVSs with large ejection velocities will escape the bulge. Only 2.6 per cent of GC HVSs are found today within $25^\circ$ of the LMC centre. The stark differences between the spatial distributions of GC HVSs and LMC HVSs lend optimism to the prospect of distinguishing the two populations.

In Fig. \ref{fig:GCv_GCdist} we show the distribution of HVS velocities and distances in the Galactocentric rest frame after each star has been propagated through the MW+LMC potential. GC-ejected stars with large ejection velocities escape the Galaxy and end up today $\sim$thousands of kpc away with Galactocentric velocities not much lower than their ejection velocities. Stars with lower ejection velocities, however, are confined to the inner few kiloparsecs of the Galaxy with significantly lower velocities. LMC HVSs show comparatively tighter distributions in Galactocentric distance and velocity, concentrated near to and slightly farther/faster than the LMC's orbital velocity ($320 \, \mathrm{km \ s^{-1}}$) and distance ($49 \, \mathrm{kpc}$). The median rest frame Galactocentric velocity for LMC HVSs is actually higher than the median ejection velocity in the LMC rest frame. Two factors contribute to this increase -- the significant orbital velocity of the LMC in the Galactocentric rest frame and the acceleration of stars ejected towards the MW as they fall deeper into the MW potential well.

\begin{figure*}
    \includegraphics[width=2\columnwidth,trim=0 0 0 170,clip]{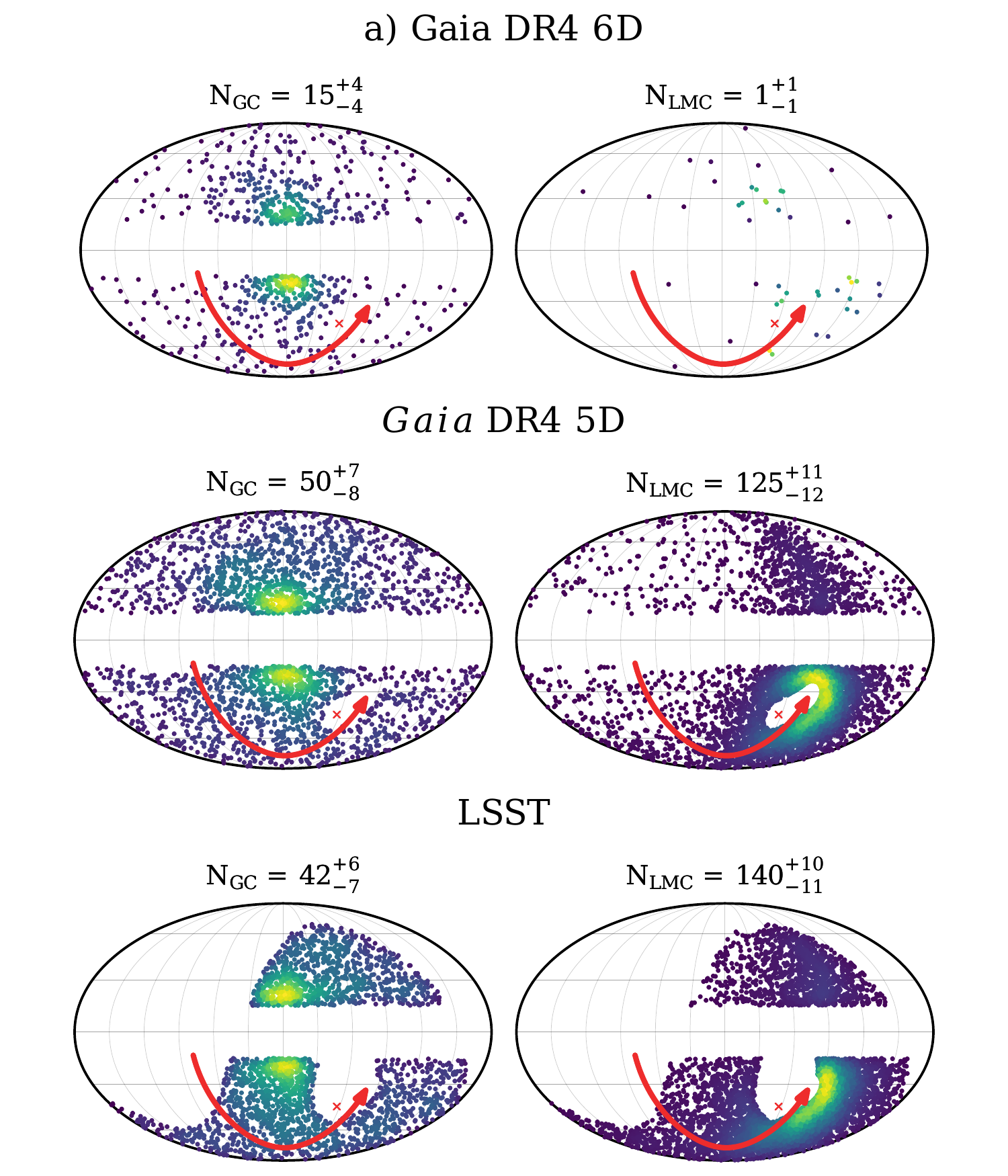}
    \caption{Sky distributions of the GC HVS and LMC HVS populations in a Mollweide projection in Galactic coordinates. \textit{Gaia} HVSs shown in upper panels, LSST HVSs in the bottom panels. See Sec. \ref{sec:methods:cuts} for more information on the sample selection criteria. Annotations show the number of detectable GC and LMC HVSs in each sample \textcolor{black}{assuming an HVS ejection rate of $10^{-4} \, \mathrm{yr^{-1}}$}. Shown distributions are stacked over 50 iterations of the simulations. Points are linearly coloured by number density The red curve shows the trajectory of the LMC on the sky over the last Gyr. The current position of the SMC is marked by a red `x' for reference.}
    \label{fig:densitydefault}
\end{figure*}

\begin{table*}
\begin{tabular}{llcccccccc}
& & $N$ & mass ($M_\odot$) & mag. & $v_{\rm GC}$ ($\mathrm{km \ s^{-1}}$) & $r_{\rm GC}$ ($\mathrm{kpc}$) & $t_{\rm age}$ (Myr) & $N_{\rm unbound}$ & $N_{\rm v>500}$ \\ \hline

                          

\multirow{2}{*}{\textit{Gaia}} & GC HVS & $50_{-8}^{+7}$ & $2.6_{-0.4}^{+1.1}$ &  $17.8_{-2.4}^{+2.0}$ & $470_{-340}^{+560}$ & $29_{-22}^{+45}$ & \cmmnt{$80_{-60}^{+180}$} $310_{-210}^{+330}$ & $25\pm6$ & $24\pm5$     \\

& LMC HVS & $125_{-12}^{+11}$  & $2.4_{-0.3}^{+0.7}$ &  $19.5_{-1.1}^{+0.8}$ & $350_{-90}^{+210}$ & $54_{-15}^{+12}$ & \cmmnt{$160_{-100}^{+90}$} $370_{-200}^{+260}$ & $40_{-8}^{+7}$ & $27\pm6$\\ \hline

\multirow{2}{*}{LSST}  & GC HVS & $42_{-7}^{+6}$  & $2.4_{-0.3}^{+0.6}$ & $20.5_{-3.1}^{+2.5}$ & $630_{-470}^{+530}$     & $64_{-56}^{+180}$ & \cmmnt{$150_{-100}^{+220}$} $420_{-240}^{+300}$& $30_{-6}^{+5}$ & $27\pm5$   \\
              
& LMC HVS & $140_{-11}^{+10}$  &  $2.3_{-0.2}^{+0.6}$ & $20.2_{-1.3}^{+1.7}$  & $300_{-80}^{+150}$   & $75_{-29}^{+73}$ & \cmmnt{$210_{-130}^{+210}$} $420_{-220}^{+250}$ & $64_{-7}^{+8}$ & $41\pm7$
\end{tabular}
\caption{Properties of our \textit{Gaia} and LSST samples of GC and LMC HVSs. The table summarizes their predicted number ($N$), median
stellar masses, apparent magnitudes, Galactocentric velocities/distances, stellar ages, the number $N_{\rm unbound}$ of HVSs gravitationally unbound to the Galaxy, and the number $N_{\rm v>500}$ of stars with Galactocentic total velocities in excess of $500 \, \mathrm{km \ s^{-1}}$. \textcolor{black}{Population counts assume an HVS ejection rate of $10^{-4} \, \mathrm{yr^{-1}}$ from both the GC and LMC.} Quoted uncertainties span the 16th to 84th percentile and are calculated over 50 iterations. \textit{Gaia} and LSST magnitudes are given in the \textit{Gaia G} and LSST \textit{r} bands, respectively.}
\label{tab:N}
\end{table*}

\section{Hyper-velocity stars observable by future surveys} \label{sec:results:cuts}

In the previous Section we explored the properties of \textit{all} surviving HVSs ejected from the GC and LMC centre. Here we select and describe stars both bright enough to be visible by \textit{Gaia} and/or LSST, and conspicuous enough to be recognized as promising HVS candidates.

\subsection{Selecting \textit{Gaia}- and LSST-detectable stars} \label{sec:methods:cuts}

In the left panel of Fig. \ref{fig:magnitudes} we show with solid lines the apparent magnitude distributions of $m>0.5 \, \mathrm{M_\odot}$ HVSs in the \textit{Gaia G} band (solid lines). For its fourth and nominally final data release (DR4), \textit{Gaia} aims to acquire five-parameter (5D) astrometric solutions as well as photometry for $\sim$2 billion sources to a depth of $G\lesssim$20.7 \citep[see][]{Gaia2016Prusti}. \textcolor{black}{We select \textit{Gaia} DR4-visible HVSs as simply all those brighter than this limit. We discuss the \textit{Gaia} selection function in more detail in Sec. \ref{sec:disc:selection}.} As seen in Fig. \ref{fig:magnitudes}, the vast majority of HVSs are fainter than $G=20.7$. An HVS with the median stellar mass among our sample ($m\simeq0.69 \, \mathrm{M_\odot}$) must be closer than $\sim 5 \, \mathrm{kpc}$ to satisfy this limit. Since more massive ejected stars are much more likely to be detected (see below for more on this point), we show as well with dash-dotted curves the distributions for GC and LMC HVSs more massive than $2 \, \mathrm{M_\odot}$, which constitute $\sim$0.75\% of the total population. The apparent magnitude of a $2 \, \mathrm{M_\odot}$ HVS at the position and distance of the LMC coincides roughly with the \textit{Gaia} DR4 magnitude limit. Two thirds of $m>2 \, \mathrm{M_\odot}$ LMC HVSs and one third of $m>2 \, \mathrm{M_\odot}$ GC HVSs are brighter than this limit. 

\textit{Gaia}'s on-board Radial Velocity Spectrometer (RVS) will provide radial velocity measurements for sources brighter than magnitude 16.2 in the \textit{Gaia} $G_{\rm RVS}$ band \citep{Katz2019}. We show the apparent magnitude distributions of $m>0.5 \, \mathrm{M_\odot}$ and $m>2 \, \mathrm{M_\odot}$ HVSs in the center panel of Fig. \ref{fig:magnitudes}. While $\sim$5 per cent of $m>2 \, \mathrm{M_\odot}$ GC HVSs will be brighter than $G_{\rm RVS}=16.2$, this cut will exclude virtually all LMC HVSs. Additionally, as of Early Data Release 3, \textit{Gaia} radial velocities are only validated for sources with effective temperatures  $\lbrack 3550 \, \mathrm{K}<T_{\rm eff}<6900 \, \mathrm{K}\rbrack$, which corresponds to a mass range of roughly $\lbrack0.5 \, \mathrm{M_\odot} \lesssim m \lesssim 2 \, \mathrm{M_\odot}\rbrack$ given our assumptions. While the final \textit{Gaia} data release will include radial velocities for hotter and cooler stars \citep{Katz2019}, the expected precision worsens at the hot end\footnote[8]{see performance predictions here: \url{https://www.cosmos.esa.int/web/gaia/science-performance}}. Given the inability of the \textit{Gaia} DR4 6D survey to detect LMC HVSs, we will not discuss it further in this paper. \textcolor{black}{GC HVSs which will appear in \textit{Gaia} DR4 6D and the constraints they can offer on the properties of the GC stellar population will be the topic of an upcoming work.}

LSST's primary `wide-deep-fast' survey strategy will observe roughly 18,000 square degrees spanning $-65^{\circ}<\delta<+5^{\circ}$ to a single-visit depth of $r<24.5$ and an estimated coadded depth of $r<27.5$ after ten years of operation \citep[see][]{Marshall2017, Ivezic2019}. We select LSST-detectable mock HVSs as those brighter than the single-visit depth in this sky volume. We remove stars brighter than LSST's saturation limit at $r=16$ \citep{Ivezic2019}. The apparent magnitude distributions of HVSs in the LSST $r$ band are shown in the right panel of Fig. \ref{fig:magnitudes}. 95 per cent of $m>2 \, \mathrm{M_\odot}$ LMC HVSs and half of $m>2 \, \mathrm{M_\odot}$ GC HVSs will satisfy $16<r<24.5$. These proportions rise to 98 per cent and 66 per cent respectively when considering $16<r<27.5$. 

We caution that an HVS satisfying a particular survey's selection criteria does not mean it will be easily \textit{identifiable} as a genuine HVS. Many stars will be ejected at low velocities and remain confined to the crowded GC or inner LMC environments (Figs. \ref{fig:vejs}, \ref{fig:skycontour}). Low-mass stars ejected to the Galactic stellar halo may be difficult to distinguish as they will likely have total velocities and spectral types typical of ordinary halo stars. In addition to the \textit{Gaia} and LSST magnitude cuts described above, we additionally apply the following ones:

\begin{itemize}
    \item Following the philosophy of the MMT Hypervelocity Star Survey \citep{Brown2006, Brown2007, Brown2009, Brown2012}, we select only early-type stars at significant Galactic latitudes, $|b|>15^{\circ}$ \citep[see also][]{Raddi2021}. Given the lack of ongoing star formation in the stellar halo, early-type stars at faint magnitudes should not be seen at high latitudes and significant distances unless they were ejected there from their place of birth.
    We define early-type stars as $m\geq 2 \, \mathrm{M_\odot}$ stars, corresponding roughly with stars of spectral type A1V or earlier \citep[e.g.][]{Pecaut2013}.
    \item To avoid confusion with ordinary bound stars in the inner LMC or Small Magellanic Cloud (SMC), we select only stars >$10^{\circ}$ from the LMC centre, >$10^{\circ}$ from the SMC centre and >$10^{\circ}$ from any point on the line connecting the LMC centre and SMC centre on the sky. At the distance of the LMC this corresponds to a projected separation of $\sim9 \, \mathrm{kpc}$.
\end{itemize}

 For brevity, in the remainder of this work we refer to our HVS samples that satisfy the above cuts as simply \textit{Gaia} HVSs and \textit{LSST} HVSs for stars within the \textit{Gaia} DR4 5D and LSST magnitude limits respectively.


\subsection{Properties of \textit{Gaia} HVS samples}


In Fig. \ref{fig:densitydefault}, we show the the sky distributions and expected numbers of our \textit{Gaia} HVS samples (upper panels). \textcolor{black}{We predict $50_{-8}^{+7}$ GC HVSs to be present in this catalogue along with $125_{-12}^{+11}$ LMC HVSs. We note that  these estimates assume an HVS ejection rate from both the GC and LMC centre of $10^{-4} \, \mathrm{yr^{-1}}$ -- they should be scaled linearly to correspond to different ejection rate assumptions, should stricter constraints on these rates become known.} While the LMC is at a larger heliocentric distance than the GC and ejects stars at a lower average velocity than the GC (see Fig. \ref{fig:vejs}), its lower central escape velocity allows for its ejected stars to escape the inner LMC easily and reach observationally accessible regions of the sky. \textcolor{black}{Therefore it is possible that LMC HVSs will greatly outnumber GC HVSs in \textit{Gaia} DR4}. The leading arm feature discussed in Sec. \ref{sec:results:allstars} is still visible, however, the most striking feature is the tight clump of LMC HVSs slightly leading the LMC orbit. We comment further on this feature later.

\begin{figure*}
    \centering
    \includegraphics[width=2\columnwidth]{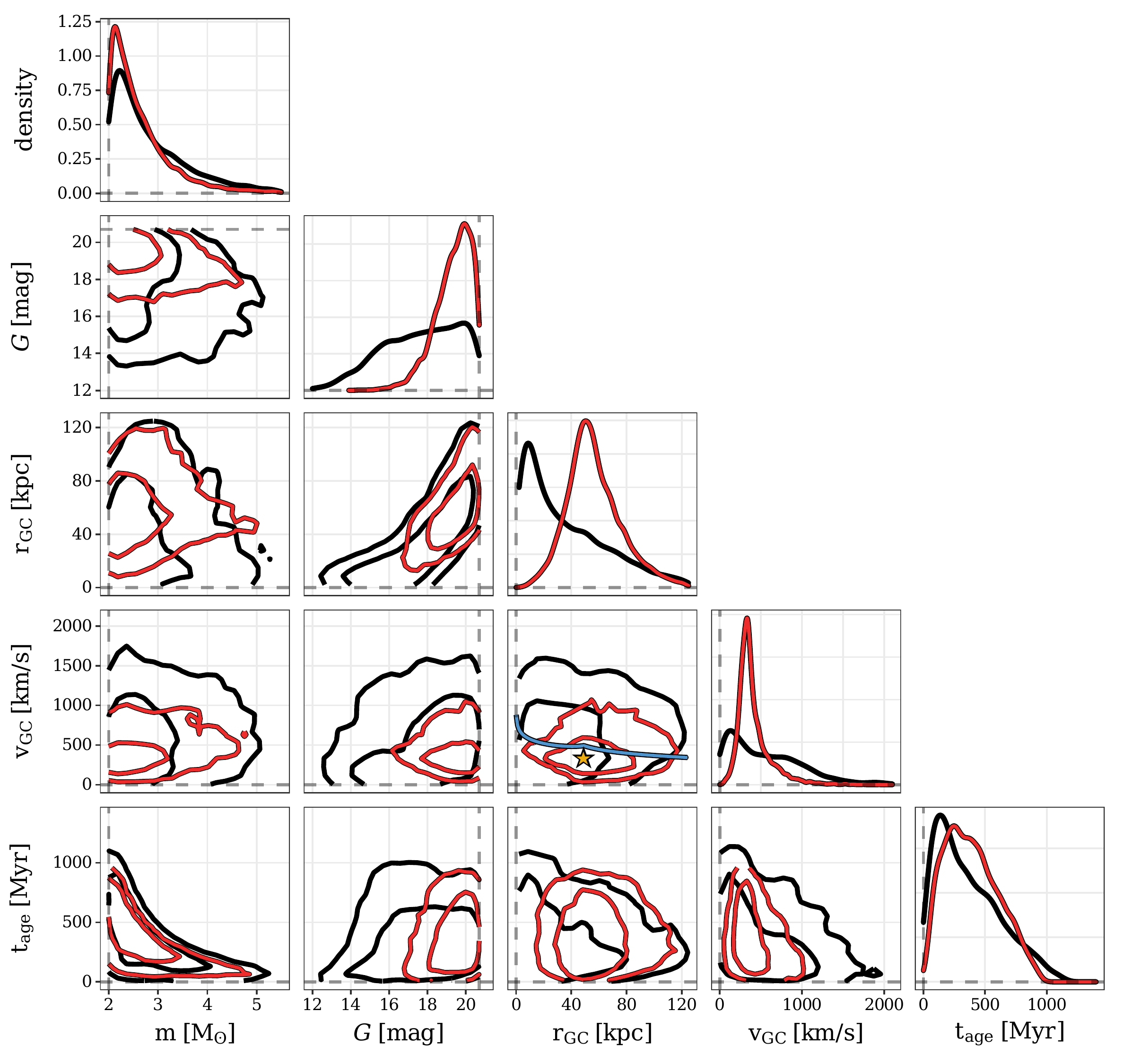}
    \caption{\textit{Gaia} HVS samples' distributions of and between stellar mass $m$, \textit{Gaia G}-band magnitude, Galactocentric distance $r_{\rm GC}$, Galactocentric total velocity $v_{\rm GC}$ and stellar age $t_{\rm age}$. LMC HVS distributions are shown in red and GC HVSs in black.
    Contours show 68th and 95th percentiles of the distributions. The blue curve in the $v_{\rm GC}$ vs. $r_{\rm GC}$ plot (fourth row, third column) shows the escape velocity to infinity from the MW+LMC potential today along the vector connecting the GC and the LMC centre. The velocity and distance of the LMC is denoted by the gold star.}
    \label{fig:stairstep}
\end{figure*}

\begin{figure*}
    \centering
    \includegraphics[width=2\columnwidth]{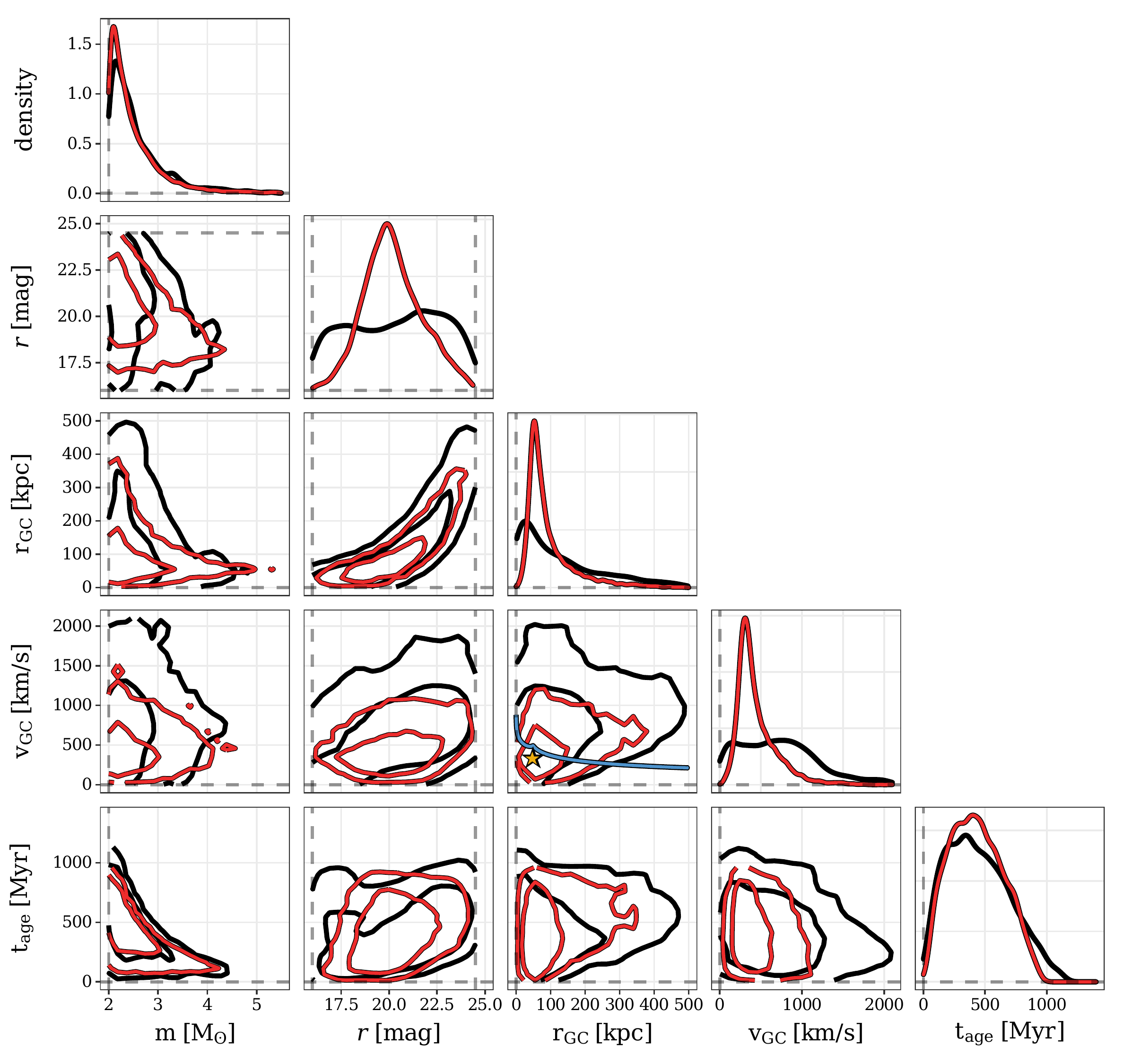}
    \caption{Same as Fig. \ref{fig:stairstep} but for the LSST samples. Apparent magnitudes are shown in the LSST $r$ band.}
    \label{fig:stairstepLSST}
\end{figure*}

We summarize the main properties of \textit{Gaia} HVSs in Table \ref{tab:N} and in Fig. \ref{fig:stairstep} we show their distributions of in stellar mass, \textit{Gaia G}-band apparent magnitude, Galactocentric distance and velocity, and stellar age. The main differences between LMC and GC HVSs are their apparent magnitudes and Galactocentric distance and velocity distributions. GC HVSs outnumber LMC HVSs by a factor of 2.2 when considering only HVSs brighter than $G=18$. While LMC HVSs are more concentrated around the Galactocentric distance and total orbital velocity of the LMC, GC HVSs have broader distributions, peaking at lower values but showing significant tails towards large distances and velocities. Roughly half ($24\pm5$) of the GC HVSs have total velocities in excess of $500 \, \mathrm{km \ s^{-1}}$, while for LMC HVSs this percentage decreases to $\sim 20\%$ $27_{-5}^{+5}$ stars).

\subsubsection{\textit{Gaia} astrometric errors}

We use \texttt{PyGaia} to characterize the estimated \textit{Gaia} end-of-mission astrometric errors of these populations. The median relative parallax error for \textit{Gaia} GC HVSs is 370 per cent. Of the $50_{-8}^{+7}$ GC HVSs predicted to appear in \textit{Gaia}, only $5\pm1$ will have a relative parallax error below 20 per cent; the relative error above which estimating distances becomes non-trivial \citep{BailerJones2015}.  LMC HVSs manage even worse, with a median relative parallax error of 1820 per cent. We do not predict a single LMC HVS to be present among our \textit{Gaia} HVSs with a relative parallax error below 20 per cent. 

By contrast, the proper motion measurements provided by \textit{Gaia} will be more reliable. The median relative uncertainty on the proper motion magnitude for GC HVSs is 6.4 per cent. $23_{-3}^{+4}$ GC HVSs will appear in \textit{Gaia} with proper motion uncertainties below 5 per cent, alike in number to the $18_{-2}^{+2}$ LMC HVSs with similar uncertainties.

\subsection{Properties of LSST HVS samples}
The spatial distribution of our LSST HVS populations are shown in Fig. \ref{fig:densitydefault}, lower panels. We predict $42_{-7}^{+6}$ GC HVSs, significantly outnumbered by the $140_{-11}^{+10}$ LMC HVSs. While LSST faint-end magnitude limit will be deeper than \textit{Gaia}, its HVS-hunting abilities are hamstrung by the fact that nearly the entirely equatorial north is inaccessible to it. If we consider LSST's ten-year estimated coadded depth of $r<27.5$ rather than its single-visit depth of $r<24.5$, we predict $142_{-12}^{+10}$ LMC HVSs and $47_{-8}^{+7}$ GC HVSs. Most $m>2 \, \mathrm{M_\odot}$ GC HVSs and nearly all $m>2 \, \mathrm{M_\odot}$ LMC HVSs already satisfy $r<24.5$ (see Fig. \ref{fig:magnitudes}), so the fainter limit does not translate to significantly more detectable HVSs. LSST's fainter limit does not imply superior astrometic errors when compared to \textit{Gaia}: \textit{Gaia} proper motion and parallax errors will be superior to LSST for sources $r\lesssim20$ and LSST smoothly extends \textit{Gaia's} error-versus-magnitude curve for $r>20$ \citep{Ivezic2012}. 

Table \ref{tab:N} and Fig. \ref{fig:stairstepLSST} summarize the properties of our LSST HVS samples. We see similar trends as in Fig. \ref{fig:stairstep}. Owing to its fainter magnitude limit, LSST GC HVSs in particular will be found at farther Galactocentric distances and faster total velocities than those found in \textit{Gaia}. LMC HVSs remain concentrated at Galactocentric total velocities similar to the LMC's orbital velocity and distances slightly farther than the LMC. For LSST, $\sim 60\%$ ($30\%$) of GC (LMC) HVSs have total velocities faster than $500 \, \mathrm{km \ s^{-1}}$, higher proportions than was the case for \textit{Gaia}.

We point out that a significant fraction of detectable HVSs will be gravitationally unbound to the MW. The blue curve in the $v_{\rm GC}$ vs. $r_{\rm GC}$ panel of Figs. \ref{fig:stairstep} and \ref{fig:stairstepLSST} (fourth row, third column) shows the escape velocity from our default MW+LMC potential along the line connecting the GC and the present-day LMC centre. Roughly half ($25\pm6$) of \textit{Gaia} GC HVSs and $\sim$one third ($40_{-8}^{+7}$) LMC HVSs will be unbound to the Galaxy. These proportions rise to 70 per cent ($30_{-6}^{+5}$) and 46 per cent ($64_{-7}^{+8}$) respectively for our LSST HVS sample (see Table \ref{tab:N}). 

\begin{figure*}
    \includegraphics[width=2\columnwidth, trim=-20 0 0 50, clip]{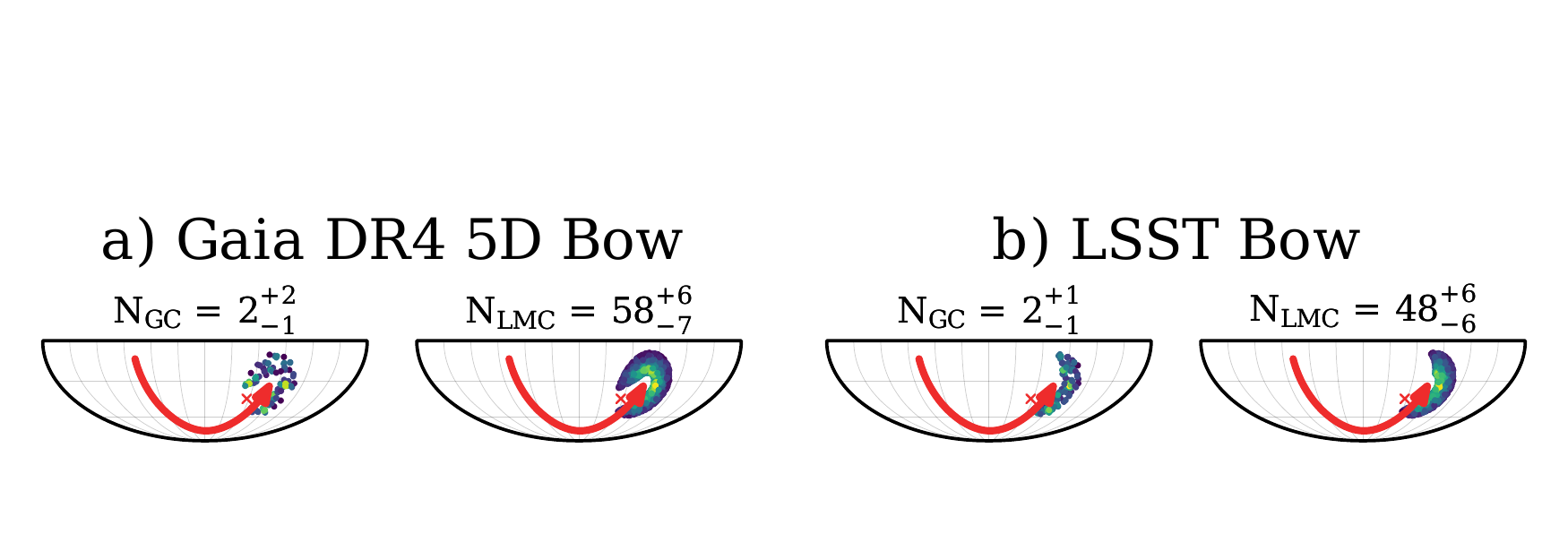}
    \includegraphics[width=2\columnwidth]{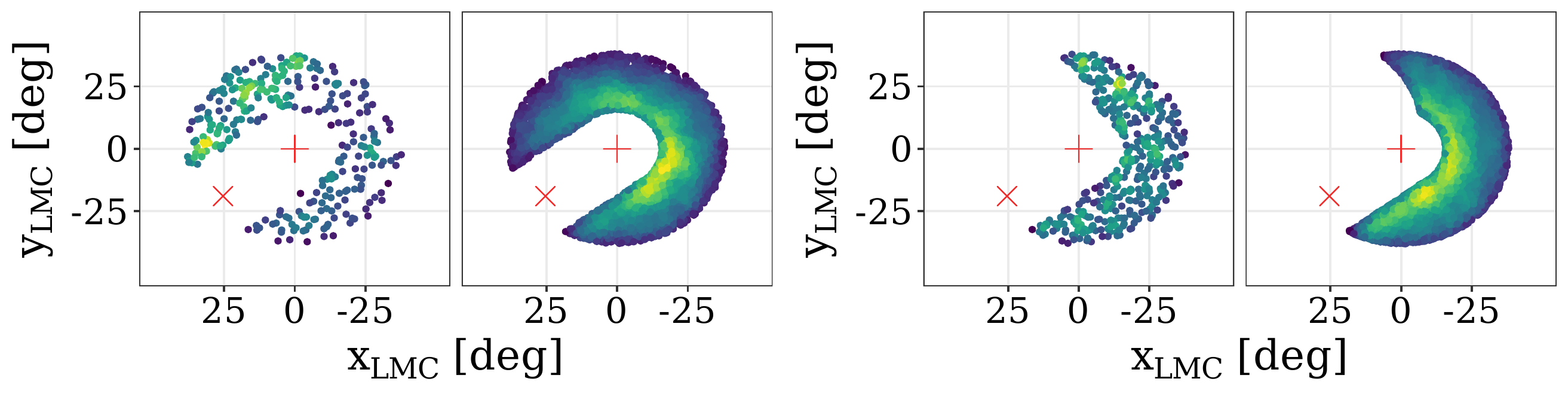}
    \caption{\textit{Top row:} Same as Fig. \ref{fig:densitydefault} but showing only GC HVSs and LMC HVSs in the bow overdensity leading the LMC orbit, i.e. >$10^\circ$ from both the LMC and SMC but <$25^\circ$ from the LMC. Distributions are shown for stars which would appear in a) \textit{Gaia} and b) LSST samples. Annotations show the number of GC and LMC HVSs present in this feature \textcolor{black}{assuming an LMC HVS ejection rate of $10^{-4} \, \mathrm{yr^{-1}}$}. \textit{Bottom row:} Same as top row but in an orthographic projection centred on the LMC centre in equatorial coordinates. Red cross and `x' symbols denote the positions of the LMC and SMC in this projection, respectively.}
    \label{fig:densitydefault_bow}
\end{figure*}

\subsection{The LMC HVS leading overdensity}
Fig. \ref{fig:densitydefault} shows that many of the LMC HVSs visible by \textit{Gaia} and LSST are found in a feature slightly leading the LMC orbit. We focus further on this overdensity by selecting only the HVSs between 10$^\circ$ and 25$^\circ$ from the LMC centre and >$10^\circ$ from the SMC centre. This $25^\circ$ limit corresponds to a projected distance of $22 \, \mathrm{kpc}$ from the LMC centre and is more or less the ballistic travel distance of a $2 \, \mathrm{M_\odot}$ HVS with the median ejection velocity from the LMC and median flight time in our fiducial model ($102 \, \mathrm{km \ s^{-1}}$ and $162 \, \mathrm{Myr}$ respectively). Note that the LSST wide-deep-fast survey strategy will not cover this entire region as it extends slightly south of the $\delta=-65^\circ$ survey edge. While stars in this bow overdensity would be sufficiently well-separated from the LMC to not be confused with inner LMC stars, they may overlap on the sky with structures seen in the LMC outskirts \citep[see e.g.][]{Gaia2020}. In particular it may overlap the stream-like feature due north of the LMC reported by \citet{Mackey2016}, though this feature is comprised predominantly of old stars.

We show in Fig. \ref{fig:densitydefault_bow} the sky distribution and number of stars located in this feature. We show as well the distribution in  an orthographic projection centred on the LMC \citep[see also][]{Gaia2018, Gaia2020}. Of the $125_{-12}^{+11}$ \textit{Gaia} LMC HVSs, $58_{-7}^{+6}$ are located in this bow overdensity feature and a similar number is expected with LSST ($48\pm6$). 
Contamination from GC HVSs in this region is modest, up to just a few \textcolor{black}{assuming an ejection rate of $10^{-4} \, \mathrm{yr^{-1}}$}. Targeting candidates in this feature in either survey for follow-up observations would therefore maximize one's chances of finding genuine LMC HVSs, while minimizing contamination from interloping GC HVSs. 

\section{Impact of MW+LMC potential assumptions} \label{sec:potentialresults}

\begin{figure*}
    \centering
    \includegraphics[width=1.5\columnwidth]{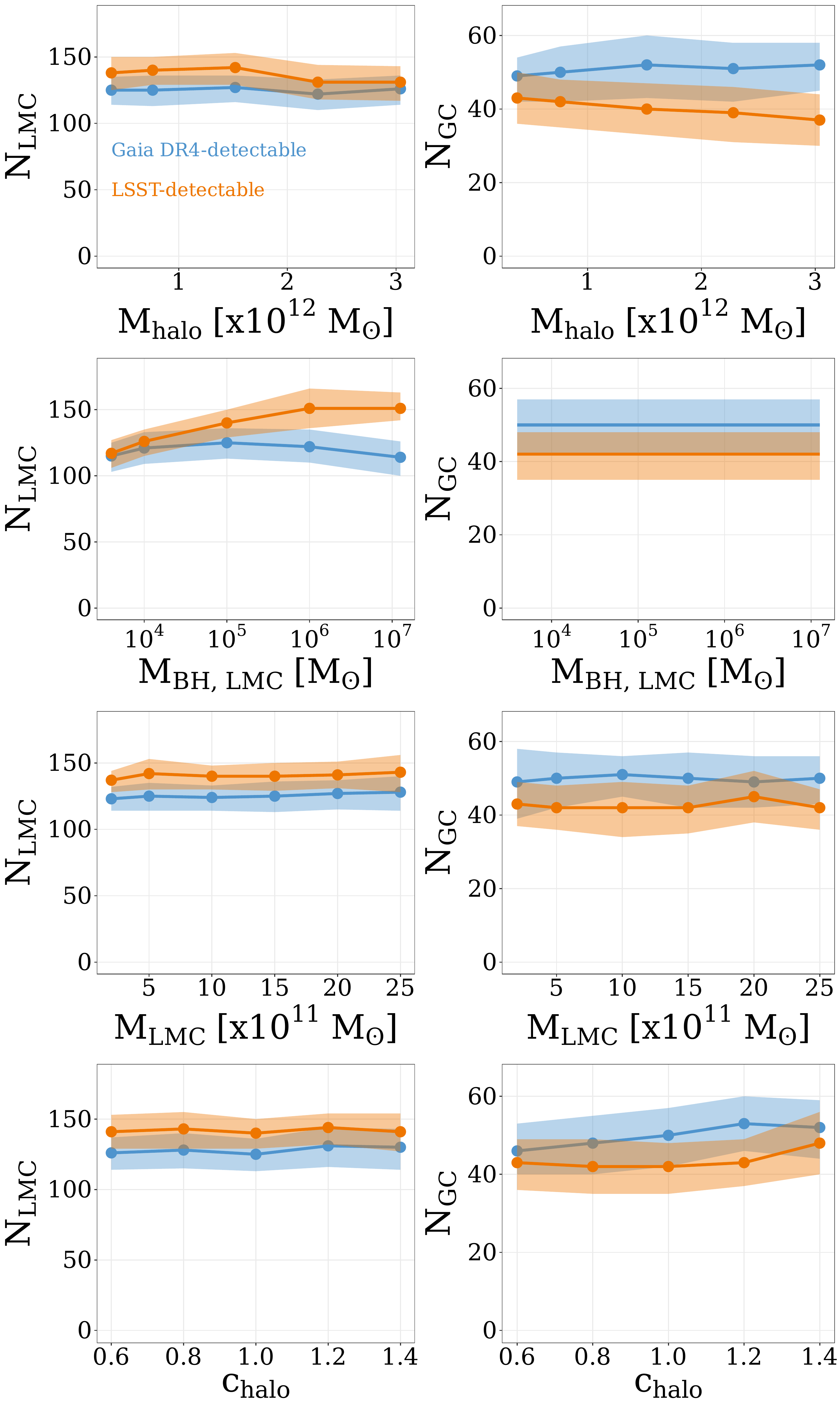}
    \caption{The dependence of the numbers $N_{\rm LMC}$ and $N_{\rm GC}$ of  \textit{Gaia}-detectable and LSST-detectable GC HVSs and LMC HVSs on model assumptions. $N_{\rm LMC}$ (left column) and $N_{\rm GC}$ (right column) are shown against the scale mass of the MW dark matter halo (top row), the mass of the LMC central MBH (second row), the LMC scale mass (third row),  MW dark matter halo flattening (fourth row). Shaded region show 1$\sigma$ confidence intervals.}
    \label{fig:NHVSParamGaia}
\end{figure*}

The maps and number estimates shown in Fig. \ref{fig:densitydefault} and Table \ref{tab:N} are for the default ejection model and MW+LMC potential summarized in Sec. \ref{sec:methods:potential}. In this subsection we run additional simulations to explore the sensitivity of these predictions to changes in our assumptions concerning the LMC MBH mass and our MW+LMC potential.

We explore additional LMC MBH masses of $M_{\rm MBH, LMC}=\lbrack10^{3.6},10^{4},10^{6},10^{7.1}\rbrack \, \mathrm{M_\odot}$. This lower limit is based on the ejection velocity of HVS3 \citep{Erkal2019} and the upper limit based on \citet{Boyce2017}, who use line-of-sight velocity maps from MUSE observations to constrain the LMC MBH mass. An increase in $M_{\rm MBH, LMC}$ affects the trajectories of LMC HVSs and leads to a larger high-velocity tail in ejection velocities (see Eq. \ref{eq:vej}). The median ejection velocity from a $M_{\rm MBH, LMC}=10^{7.1} \, \mathrm{M_\odot}$ MBH increases to $440 \, \mathrm{km \ s^{-1}}$ compared to the median ejection velocity of $200 \, \mathrm{km \ s^{-1}}$ from our default $M_{\rm MBH, LMC}=10^{5} \, \mathrm{M_\odot}$ MBH. 

When considering different LMC masses, we explore $M_{\rm LMC}=\lbrack 2, 5, 10, 20, 25\rbrack \times 10^{10} \, \mathrm{M_\odot}$ similar to \citet{Belokurov2019}. The lower limit is based on the enclosed mass constraints within $8.7 \, \mathrm{kpc}$ inferred by \citet{vanderMarel2014} via rotation measurements. The upper limit is based on the LMC mass determined by \citet{Penarrubia2016} based on the impact of the LMC on the MW-M31 timing argument.  We set a different LMC scale radius for each scale mass, requiring that the circular velocity at $8.7 \, \mathrm{kpc}$ from the LMC centre is $91.7 \, \mathrm{km\ s^{-1}}$ \citep{vanderMarel2014}. 

While the Galactic disc and bulge potentials remain fixed in all models, we consider changes in the Galactic dark matter halo. We explore spherical NFW halos ($c_{\rm halo}$=1) with $r_{\rm s}$ fixed and $M_{\rm s}$=$\lbrack0.5,2,3,4\rbrack \times0.76\times10^{12} \, \mathrm{M_{\odot}}$, as well as oblate/prolate spheroids with $M_{\rm s}$ and $r_{\rm s}$ fixed at their fiducial values but $c_{\rm halo}$=$\lbrack0.6,0.8,1.2,1.4\rbrack$.

In Fig. \ref{fig:NHVSParamGaia} we show how the numbers of \textit{Gaia} and LSST HVSs depend on the the above parameters. Both the predicted number of GC HVSs ($N_{\rm GC}$) and LMC HVSs ($N_{\rm LMC}$) are fairly agnostic towards $M_{\rm halo}$. The number of LSST HVSs ejected from both the GC and LMC declines slightly with increasing $M_{\rm halo}$. A heavier halo implies a larger escape velocity from the inner Milky Way and therefore a reduced number of GC HVSs reaching observationally accessible locations on the sky. However, a larger deceleration from the GC to the Galactic halo reduces the typical heliocentric distance of GC HVSs at time of observation, increasing the number of brighter sources within the \textit{Gaia} or LSST  horizons. As $M_{\rm halo}$ increases, the LMC has spent a larger fraction of its recent past close to the MW, so stars with larger flight times become bright enough to be detected. Additionally, as with $N_{\rm GC}$, a larger enclosed mass implies fewer LMC HVSs on outbound orbits reach Galactocentric distances beyond the \textit{Gaia} horizon.

We see in Fig. \ref{fig:NHVSParamGaia} that increasing $M_{\rm BH, LMC}$ increases the number of LSST LMC HVSs. The number of \textit{Gaia} LMC HVSs increases mildly with $M_{\rm BH, LMC}$, peaking near $M_{\rm BH, LMC}\simeq 10^5 \, \mathrm{M_\odot}$ before turning over. Recall from Eq. \ref{eq:vej} that the Hills mechanism ejection velocity scales as $\propto M_{\rm BH}^{1/6}$ -- an order of magnitude increase in $M_{\rm BH}$ scales $v_{\rm ej}$ by a factor of $\sim 1.5$. A more massive LMC MBH is more effective at ejecting stars which leave the inner LMC, but simultaneously ejects more stars which reach heliocentric distances beyond the \textit{Gaia}/LSST horizons before being observed.

 We see in Fig. \ref{fig:NHVSParamGaia} that $N_{\rm GC}$ and $N_{\rm LMC}$ are quite flat with  $M_{\rm LMC}$. In the $M_{\rm LMC}$ range we probe, the LMC remains far sub-dominant to the MW in its contribution to the MW+LMC potential. A modest increase in $M_{\rm LMC}$ does not appreciably affect the fraction of LMC HVSs which escape the inner LMC. While the LMC mass impacts the perturbation of GC HVSs by the LMC \citep[see][]{Kenyon2018}, its impact on the absolute number of observable GC HVSs is negligible. We note that even when the LMC scale radius is kept fixed, $N_{\rm GC}$ and $N_{\rm LMC}$ do not vary with increasing $M_{\rm LMC}$.

Finally, we show in Fig. \ref{fig:NHVSParamGaia} the impact of the flattening of the Galactic dark matter halo. $c_{\rm halo}<1$ and $c_{\rm halo}>1$ correspond to an oblate and a prolate halo, respectively. While $N_{\rm LMC}$ is quite flat with $c_{\rm halo}$, $N_{\rm GC}$ overall increases mildly with increasing $c_{\rm halo}$ for both \textit{Gaia} and LSST HVS populations. While different halo flattenings affect the trajectories of individual stars and the orbital history of the LMC, its impact on the number of detectable HVSs is quite small. The deflection caused by an oblate halo may bring more GC HVSs towards the Galactic midplane and thus not satisfy our $|b|>15^\circ$ cut. This is the most likely explanation for the slight increase of $N_{\rm GC}$ with $c_{\rm halo}$.

In summary, while the model parameters explored above impact the trajectories and detectability of individual stars, the absolute numbers of GC HVSs and LMC HVSs observable by \textit{Gaia} and LSST are relatively robust against variation of these parameters. Marginalizing over these explored parameters in the ranges explored assuming flat priors for $M_{\rm halo}$, $c_{\rm halo}$, $M_{\rm LMC}$ and a log-flat prior for $M_{\rm BH, LMC}$, we obtain $N_{\rm GC}$ and $N_{\rm LMC}$ predictions quite close to the estimates of our fiducial model.

\section{Discussion}
\label{sec:discussion}
In this section we comment on a number of considerations related to our predictions and on caveats which may impact the expected number and characteristics of GC and LMC HVSs. 

\subsection{Comparison with known HVS candidates}

 HVS3 was discovered by \cite{Edelmann2005} while performing a spectroscopic follow-up of faint B-type star candidates from the Hamburg/ESO survey. This is the only HVS candidate currently associated with an origin in the LMC centre \citep{Erkal2019}. It appears in \textit{Gaia} DR2 with an apparent magnitude of $G=16.4$, but would not have been kinematically identified as a promising HVS candidate from \textit{Gaia} data alone, as it lacks a reliable parallax measurement and \textcolor{black}{does not appear in the \textit{Gaia} subsample with published radial velocities.} 
 
 HVS3 is a $8\pm1 \, \mathrm{M_\odot}$ star located at $\alpha=69.6^\circ$, $\delta=-54.6^\circ$ \citep{Przybilla2008HVS3}. This places is coincident with the bow overdensity feature leading the LMC orbit shown in Fig. \ref{fig:densitydefault_bow}. While its distance and total velocity are typical of other $m>5 \, \mathrm{M_\odot}$ stars located in this feature, according to our results it is quite a rare and exceptional object. \textcolor{black}{At an assumed LMC HVS ejection rate of $10^{-4} \, \mathrm{yr^{-1}}$, we} predict that \textit{Gaia} DR4 will contain $1.5_{-0.9}^{+1.4}$ LMC HVSs with masses greater than $5 \, \mathrm{M_\odot}$ in this bow feature, and we find a 6 per cent chance for it to contain a $m\geq8 \, \mathrm{M_\odot}$ star.  While not incompatible with our predictions, the presence of HVS3 in \textit{Gaia} could hint that our LMC HVS ejection rate assumption of $10^{-4} \ \mathrm{yr^{-1}}$ errs to the conservative side. \textcolor{black}{If the true LMC HVS ejection rate is a factor of $\sim$a few larger than assumed here, the existence of an HVS3-like object becomes much less unlikely.} Alternatively, the initial mass function in the LMC centre may be more top-heavy than assumed in this work. \textcolor{black}{For instance, the probability of \textit{Gaia} DR4 containing an HVS3-like object in the leading overdensity doubles to $\sim12$ per cent if the stellar IMF in the innermost pc of the LMC has a power-law slope of $-1.7$ similar to the IMF in the innermost $0.5 \, \mathrm{pc}$ of the Milky Way found by \citet{Lu2013}}.
 
Equally exceptional as HVS3 is the confirmed GC HVS S5-HVS1 \citep{Koposov2020}. While first  identified in the S$^5$ survey \citep{Li2019}, S5-HVS1 appears in \textit{Gaia} DR2 with an apparent magnitude of $G=16.0$ and without a reliable parallax measurement or radial velocity measurement. At $\alpha=343.7^\circ$, $\delta=-51.2^\circ$, it is located just south of the GC HVS overdensity in the Galactic south seen in Fig. \ref{fig:densitydefault}. S5-HVS1 has a breakneck estimated ejection velocity ($v_{\rm ej}\simeq \, 1800 \, \mathrm{km \ s^{-1}}$) and relatively short flight time ($t_{\rm flight}\simeq 4.8 \, \mathrm{Myr}$). We do not predict future \textit{Gaia} data releases to contain a sizeable population of S5-HVS1-like objects -- we expect \textit{Gaia} DR4 to contain only $1.6_{-0.9}^{+1.4}$ GC HVSs with $v_{\rm ej}>1500 \, \mathrm{km \ s^{-1}}$ and $t_{\rm flight}<20 \, \mathrm{Myr}$.

We finally consider the HVS candidates in the MMT HVS Survey \citep{Brown2012}, gathered from a targeted search of early-type stars in the Galactic halo. In particular, we consider the 12 HVS candidates whose distribution of possible Galactic plane-crossing locations include the GC at the 1$\sigma$ level \citep[c.f.][fig. 8]{Kreuzer2020}. Like HVS3 and S5-HVS1, these objects all appear in \textit{Gaia} DR2 without radial velocities. They are quite distant ($50 \, \mathrm{kpc}\lesssim d \lesssim 160 \, \mathrm{kpc}$) -- with such long flight times and large relative astrometric uncertainties, the range of possible plane-crossing locations becomes naturally quite extended, often many times larger than the size of the Galactic disc. Because our expected GC HVS sample peaks at distances much smaller than 50 kpc (see Table \ref{tab:N} and Fig. \ref{fig:stairstep}), these 12 known candidates would constitute a substantial fraction of the $15_{-3}^{+4}$ GC HVSs we predict to appear in \textit{Gaia} DR4 at heliocentric distances in excess of $50 \, \mathrm{kpc}$.

\subsection{HVS ejection rates from the GC and LMC centre} \label{sec:disc:rates}

\begin{figure}
    \includegraphics[width =0.85\columnwidth,trim=0 0 0 0,clip]{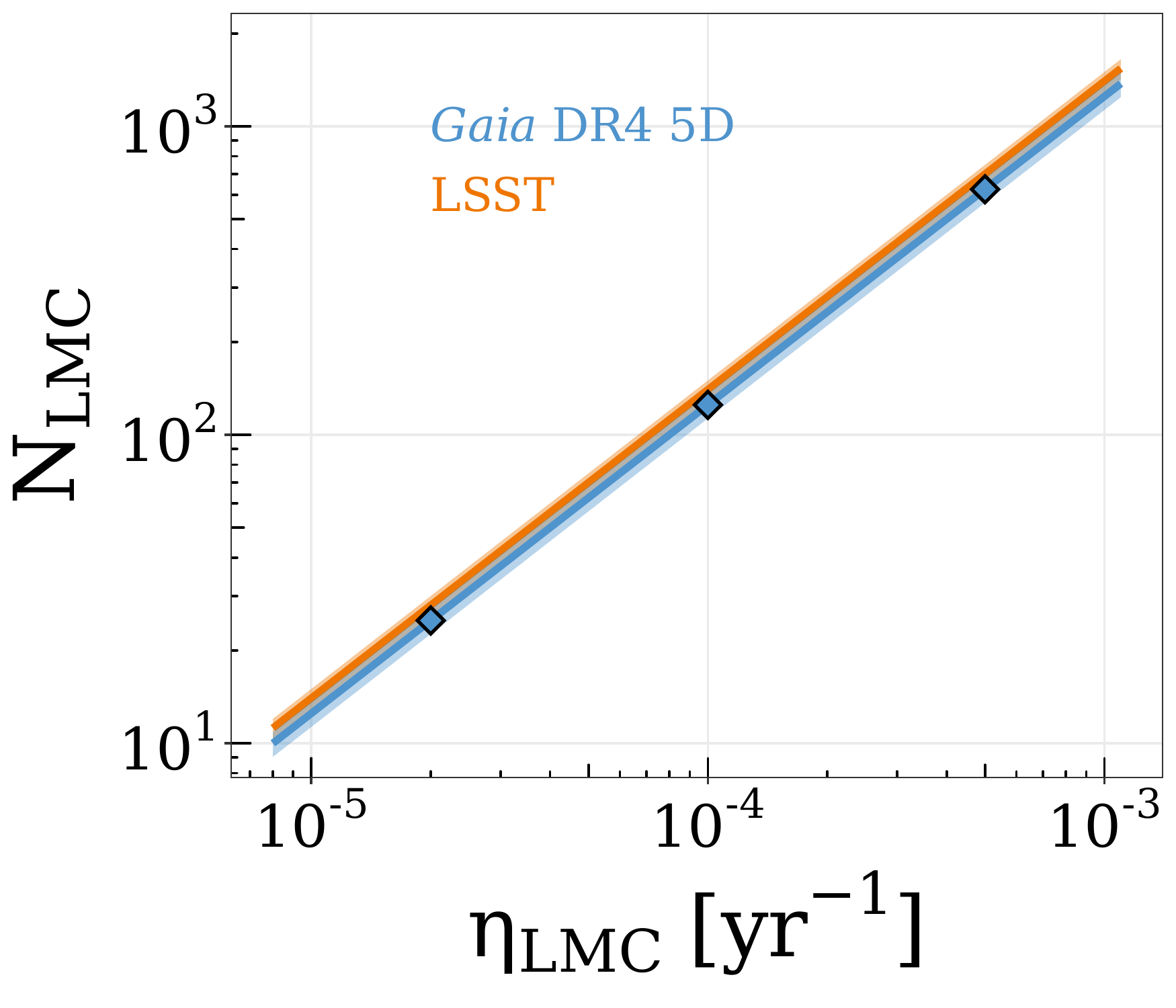}
    \begin{center}
    \includegraphics[width =\columnwidth]{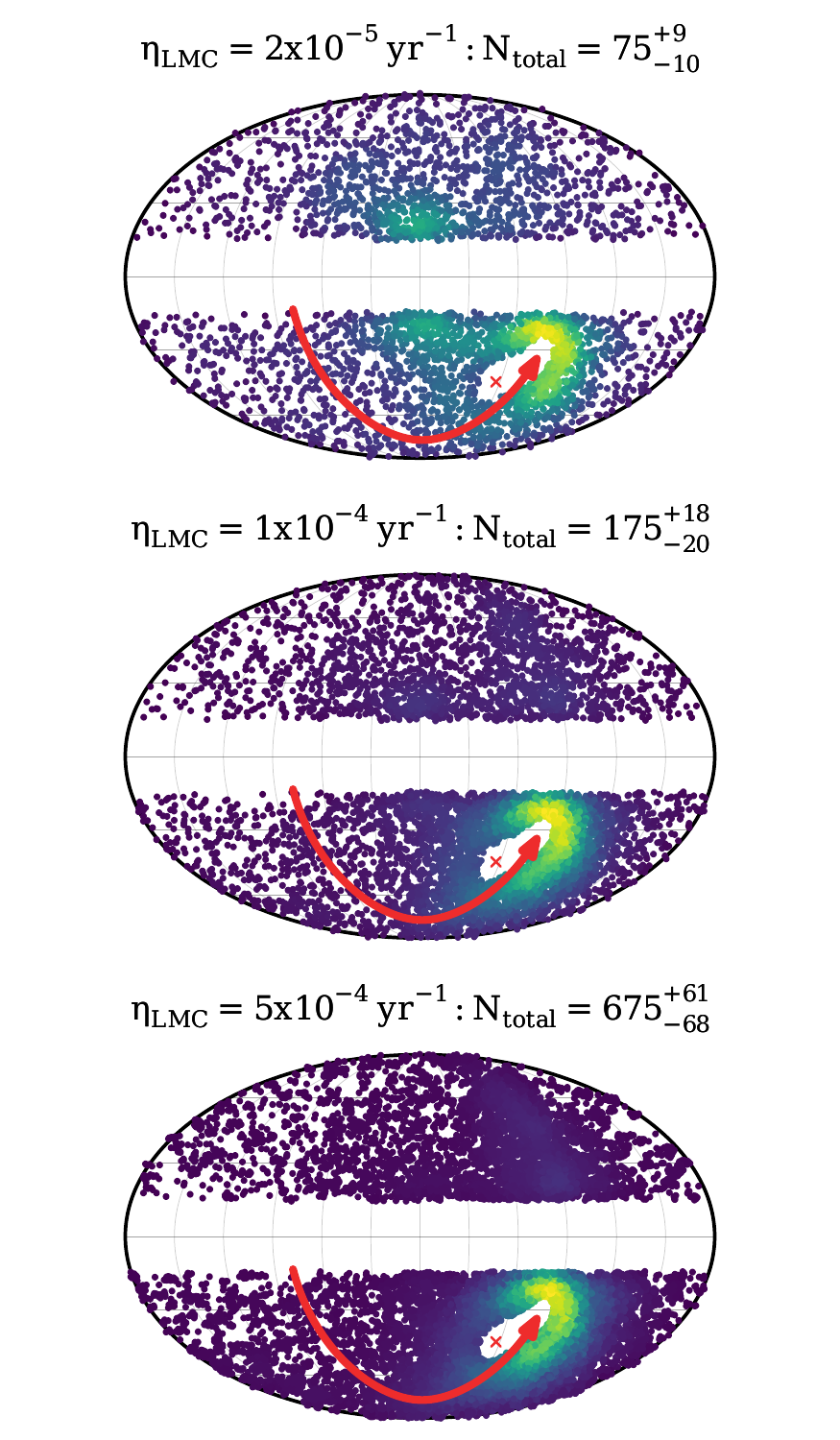}
    \caption{\textcolor{black}{\textit{Top}: Dependence of $N_{\rm LMC}$ on the LMC HVS ejection rate $\eta_{\rm LMC}$ for HVSs detectable by both \textit{Gaia} and LSST. \textit{Second, third and fourth from top:} Similar to Fig. \ref{fig:densitydefault} but for the combined GC+LMC HVS population detectable in \textit{Gaia} DR4. Different $\eta_{\rm LMC}$ values are assumed in each panel corresponding to the plotted points in the top panel, while the GC HVS ejection rate is fixed at $10^{-4} \, \mathrm{yr^{-1}}$.}}
    \label{fig:skydensityeta}
    \end{center}
\end{figure}

HVS ejection rates are derived and constrained from different angles: theoretical estimates \citep[e.g.][]{Hills1988, Yu2003}, 
detailed simulations \citep{Zhang2013}, comparisons to known HVS candidates \citep[e.g.][]{Bromley2012,Brown2014, Marchetti2018}, observations of the Galactic S star cluster \citep{Bromley2012} and tidal disruption event rates in the local Universe \citep[e.g.][]{Wang2004, Gezari2009, Bromley2012, vanVelzen2014}. Overall, these methods indicate an HVS ejection rate in the range $\sim10^{-5}-10^{-3} \, \mathrm{yr^{-1}}$. For simplicity, in this work we adopt a constant HVS ejection rate of $10^{-4} \, \mathrm{yr^{-1}}$ for both the Milky Way and the LMC, so that a sample comparison is more straightforward. However, it might not be a realistic description. For example, the HVS ejection rate may scale inversely with the mass of the central black hole as the tidal disruption event rate does \citep[see][]{Stone2016}. In this case the HVS ejection rate in the LMC would be greater than in the GC, further increasing the predicted dominance of LMC HVSs over GC HVSs in \textit{Gaia} and LSST catalogues. 

\textcolor{black}{We demonstrate the importance of the still-uncertain LMC HVS ejection rate $\eta_{\rm LMC}$ in Fig. \ref{fig:skydensityeta}. In the top panel we illustrate how $N_{\rm LMC}$ for both \textit{Gaia} and LSST HVSs change linearly with $\eta_{\rm LMC}$. By comparison with Fig. \ref{fig:NHVSParamGaia}, it is clear that $\eta_{\rm LMC}$ impacts our predictions more strongly than any assumption regarding the LMC MBH mass or the MW+LMC potential. In the remaining panels of Fig. \ref{fig:skydensityeta} we show the sky distribution of the combined \textit{Gaia} GC and LMC HVS population for different $\eta_{\rm LMC}$ assumptions while the GC HVS ejection rate is fixed at $10^{-4} \, \mathrm{yr^{-1}}$. It is clear that a significant change in $\eta_{\rm LMC}$ greatly influences the size of the total HVS population and its distribution. If stricter constraints on the LMC HVS ejection rate become available, our predictions can be scaled accordingly. However, our characterizations of the spatial distributions of the individual LMC HVS and GC HVS populations and their kinematics and properties remain valid as they are independent of the assumed HVS ejection rate.}

\subsection{\textcolor{black}{The \textit{Gaia} and LSST selection functions}} \label{sec:disc:selection}

\textcolor{black}{Throughout this work we have assumed that the \textit{Gaia} DR4 catalogue with five-parameter astrometry will be 100 per cent complete for $G<20.7$ and entirely incomplete for fainter magnitudes, and similarly for LSST in the range $16<r<24.5$ in its survey volume. Though the precise selection functions of these future surveys cannot be predicted exactly, one may worry that the above assumptions are overly simplistic. For instance, because the \textit{Gaia} astrometric pipeline requires a source be detected a minimum number of times to be included in the catalogue \citep{Lindegren2018, Lindegren2021}, the \textit{Gaia} completeness function depends on the spinning-and-precessing scanning law of the satellite \citep[see][]{Boubert2020cogi}. \textit{Gaia} completeness is also impacted in regions of high crowding \citep{Boubert2020cogii}, however, this is not worrisome in the context of our results since we omit the most crowded regions of the sky (i.e. the inner LMC and Galactic midplane) from our analysis.}

\textcolor{black}{We consider the validity of our \textit{Gaia} DR4 selection function approximation by comparing it to a more robust approximation, leveraging the known properties of the \textit{Gaia} EDR3 selection function. We assume the following:}

\textcolor{black}{
\begin{itemize}
    \item The $G\leq21$ and \textsc{visibility\_periods\_used} $\geq 9$ requirements for a source to have a full five-parameter astrometric solution\footnote[9]{\textcolor{black}{A \textit{Gaia} visibility period is a group of astrometric observations separated from other groups by a time period of at least four days.} } in \textit{Gaia} EDR3 \citep{Lindegren2021} remain unchanged for the DR4 astrometric pipeline.
    \item The \textit{Gaia} DR4 source catalogue is not substantially more complete than EDR3 for $G\leq21$, i.e. DR4 will not contain many sources which do not already appear in EDR3 with at least a two-parameter astrometric solution.
    \item Given that DR4 will be based on observations spanning a time period double that of EDR3, nearly all $G\leq21$ sources in EDR3 will satisfy \textsc{visibility\_periods\_used} $\geq 9$ in DR4 \citep[c.f. fig. 3][]{Fabricius2021}.   
\end{itemize}
}

\textcolor{black}{If the above assumptions hold, the selection function for the \textit{Gaia} DR4 subsample with five-parameter astrometry can be well-approximated by the $G\leq21$ selection function of the \textit{Gaia} EDR3 source catalogue. If we apply the EDR3 selection function provided by the \texttt{PYTHON} package \textsc{selectionfunctions}\footnote[10]{\textcolor{black}{https://github.com/gaiaverse/selectionfunctions}} based on \citet{Boubert2020cogii} to our mock HVS populations, we end up with only 6\% more GC HVSs and 10\% more LMC HVSs in \textit{Gaia} DR4. For the purposes of this work, our simple $G<20.7$ cut can therefore be considered a reasonable and conservative approximation to the \textit{Gaia} DR4 astrometric selection function, missing the small fraction of HVSs fainter than the nominal completeness limit.}

\subsection{The reflex motion of the Milky Way in response to the LMC}

 Throughout this work we have assumed a reference frame in which the MW is fixed and the LMC orbits around it. With a mass $\sim$one tenth the mass of the MW, the LMC not only deflects the orbits of GC HVSs \citep{Kenyon2018} but additionally perturbs the MW and induces a reflex motion of the MW centre of mass. \citet{Gomez2015} show that as a result of the LMC, the MW centre of mass within $50 \,\mathrm{kpc}$ has been displaced by as much as $30 \, \mathrm{kpc}$ and $75 \, \mathrm{km \ s^{-1}}$ over the last $\sim0.3-0.5 \, \mathrm{Gyr}$. Among other impacts, this reflex motion will affect the long-term orbital history of the LMC, the tidal disruption of the Sagittarius dwarf galaxy \citep{Gomez2015}, and the apparent motions within the outer stellar halo \citep{GaravitoCamargo2019, Peterson2020,Erkal2020}. 

In the context of HVSs, \citet{Boubert2020} study the influence of this reflex motion on GC HVS trajectories. They find this reflex motion deflects the apparent trajectories of HVSs on the same order as the deflection induced by the LMC's influence itself. Disentangling these two effects, however, requires proper motion measurements precise to $10 \, \mathrm{\mu as \ yr^{-1}}$, which would be possible in the proposed \textit{Gaia} successor mission \textit{Gaia}NIR \citep{Hobbs2016}.

Due to this reflex motion, it may be more difficult to conclusively associate GC HVSs with a GC origin, as their past trajectories will reflect the fact that they were ejected from the past GC location rather than the location of the GC today. Despite this, since most GC and LMC HVSs have flight times within a few hundred Myr and Galactocentric distances within a $\sim$few tens of kpc, the number of GC-ejected and LMC-ejected \textit{Gaia} HVSs or LSST HVSs will not be significantly impacted by this phenomenon.

\subsection{Contamination by LMC hyper-runaways} \label{sec:disc:runaways}

Interaction with an MBH is not the only mechanism by which stars can obtain significant peculiar velocities. Dynamical ejections from young, dense stellar clusters \citep[see][]{Poveda1967, Leonard1990, Leonard1991} and binary disruptions following core-collapse supernovae \citep[e.g][]{Blaauw1961, Tauris1998, Renzo2019} are often blamed for the known population of main sequence `runaway stars' with ejection velocities $\geq 30-40\,\mathrm{km\ s^{-1}}$
\citep{Blaauw1961}. Both mechanisms are known, however, to struggle to eject enough stars at $\geq 450-500\,\mathrm{km\ s^{-1}}$ to explain the known population of main sequence 'hyper-runaway' stars ejected from the Galactic disc \citep{Portegies2000, Perets2012, Oh2016, Evans2020}.

With its lower escape speed and its significant orbital velocity, however, the LMC disc might be more effective at ejecting main sequence runaway stars which escape the inner LMC. \citet{Boubert2017} explore the prospect of LMC hyper-runaways ejected via binary supernovae \textcolor{black}{using the binary evolution code \texttt{binary\_c} \citep{Izzard2009} and an N-body MW+LMC potential}. Their results indicate that at least as many early-type LMC hyper-runaways as HVSs may escape from the inner parts of the LMC and be detectable by \textit{Gaia}. It may be difficult, then, to distinguish LMC HVSs from hyper-runaway stars ejected from the LMC disc. A quantitative assessment of this is beyond the scope of this paper and is deferred to a future dedicated work.

\section{Summary and Conclusion} 
\label{sec:conclusions}
In this work we perform a suite of simulations ejecting hyper-velocity stars from the centres of the MW and the LMC via dynamical interactions between their central massive black holes and nearby binary systems. By propagating these stars through the Galactic potential and obtaining mock observations, we predict and characterize the GC HVS and LMC HVS populations visible in the near future by the \textit{Gaia} mission in its fourth and final data release (DR4) as well as by the LSST survey. We restrict our predictions to stars which would be conspicuous as promising HVS candidates, i.e. early-type ($m>2 \, \mathrm{M_\odot}$) stars well-separated on the sky from the Magellanic Clouds and the Galactic midplane. Our findings are summarized as follows:
\begin{itemize}
    \item \textcolor{black}{While the HVS ejection rate from the LMC remains uncertain, we predict $125_{-12}^{+11}$ LMC HVSs will be included in \textit{Gaia} DR4 with five-parameter astrometric solutions assuming an HVS ejection rate of $10^{-4} \, \mathrm{yr^{-1}}$. These will greatly outnumber the $50_{-8}^{+7}$ GC HVSs we expect to be present in \textit{Gaia} DR4.} One half of GC HVSs and 22 per cent of LMC HVSs will have total velocities in excess of $500 \, \mathrm{km \ s^{-1}}$ (see Table \ref{tab:N} and Fig. \ref{fig:stairstep}).
    \item We do not predict any LMC HVSs to be present in the \textit{Gaia} DR4 radial velocity catalogue due to its shallower magnitude limit ($G_{\rm RVS}<16.2$) and the restrictions on stellar effective temperature in the \textit{Gaia} spectroscopic pipeline.
    \item Within LSST's wide-deep-fast survey volume there will be $42_{-7}^{+6}$ GC HVSs and $141_{-11}^{+10}$ LMC HVSs within its magnitude limits. Median Galactocentric total velocities for these objects will be $630_{-470}^{+530} \, \mathrm{km \ s^{-1}}$ and  $300_{-80}^{+150} \, \mathrm{km \ s^{-1}}$, respectively. 64 per cent of GC HVSs and 29 per cent of LMC HVSs will have total velocities faster than $500 \, \mathrm{km \ s^{-1}}$ (see Table \ref{tab:N} and Fig. \ref{fig:stairstepLSST}).
    \item GC HVSs and LMC HVSs will differ starkly in their spatial distribution (Figs. \ref{fig:skycontour} and \ref{fig:densitydefault}), concentrated mainly towards the Galactic meridian and outer LMC regions, respectively. LMC HVSs in particular will be preferentially found in a feature $<25^\circ$ from the LMC leading its orbit (Fig. \ref{fig:densitydefault_bow}). $\sim$one half of \textit{Gaia}-detectable LMC HVSs and one third of LSST-detectable LMC HVSs will be found in this feature.
    \item The above estimates are robust against variations in assumptions concerning the mass of the LMC central MBH and the MW+LMC gravitational potential (Fig. \ref{fig:NHVSParamGaia}).
\end{itemize}

With these results we show that the LMC may be a more bountiful wellspring of HVSs than was previously appreciated. The prospects are promising for current and near-future deep and wide Galactic surveys to increase the population of known HVSs significantly from both the GC and LMC. The confirmation and further study of HVSs can be used as probes to study the structure of the MW and LMC, the interactions between them, and the environments of their MBHs. Our results outline expectations for the number of HVSs we should expect to detect in near-future surveys, provide clues on how to search for these objects effectively and how to offer predictions on how GC HVSs and LMC HVSs will compare and contrast.

\section*{Acknowledgements}

\textcolor{black}{We thank the anonymous referee, whose helpful and considered feedback improved this manuscript.} We thank Omar Contigiani and Anthony Brown for helpful discussions. FAE acknowledges funding support from the Natural Sciences and Engineering Research Council of Canada (NSERC) Postgraduate Scholarship. TM acknowledges an ESO fellowship. 

\section*{Data Availability}

The simulation outputs underpinning this work can be shared upon request to the corresponding author.

\bibliographystyle{mnras}
\bibliography{main}

\bsp	
\label{lastpage}
\end{document}